\documentclass[12pt]{article}

\usepackage{geometry}
\usepackage{booktabs}
\usepackage{float}
\usepackage{graphicx} 
\usepackage{amssymb, amsmath, amsthm}
\usepackage{microtype}
\usepackage{setspace}

\usepackage{pgfplots}
\pgfplotsset{compat=1.18}

\usepackage{tikz}
\usetikzlibrary{arrows.meta,positioning,fit}
\usetikzlibrary{decorations.pathreplacing}

\usepackage{xcolor}
\definecolor{darkred}{RGB}{141, 0, 0}
\usepackage{hyperref}
\hypersetup{
    colorlinks,
    linkcolor={darkred},
    citecolor={darkred},
    urlcolor={darkred}
}

\usepackage{soul}

\newtheorem{proposition}{Proposition}
\newtheorem{theorem}{Theorem}
\newtheorem{lemma}{Lemma}

\usepackage[hang,flushmargin]{footmisc}

\usepackage{parskip}

\usepackage[style=authoryear,texencoding=utf8,backend=biber, minnames = 1, maxnames = 3]{biblatex}
\DeclareMathOperator*{\argmin}{argmin}
\DeclareMathOperator*{\argmax}{argmax}

\newcommand{\cl}{u}
\newcommand{\pl}{\ell}
\newcommand{\ny}{m}
\newcommand{\na}{n}

\long\def\private#1{\relax}  %% <-- HIDE private stuff
\title{
Misaligned by Design:\\ Incentive Failures in Machine Learning\thanks{We thank the Sloan Foundation for its support under the ``Cognitive Economics at Work'' grant. We thank Annie Liang, Preston McAfee, Ashesh Rambachan, Larry Samuelson, Jakub Steiner, Eva Tardos, and participants in the USC Conference on AI, Economics, and Business and the SMART symposium at the University of Zurich for valuable comments.}
}

\author{{David Autor}\thanks{Massachusetts Institute of Technology, Google Technology and Society Fellows program, and NBER.}, {  Andrew Caplin}\thanks{New York University and NBER.}, {  Daniel Martin}\thanks{University of California, Santa Barbara.}, {  Philip Marx}\thanks{Louisiana State University.}}
\date{November 11, 2025}

\begin{document}

\maketitle

\thispagestyle{empty}

\begin{abstract}

\noindent The cost of error in many high-stakes settings is asymmetric: misdiagnosing pneumonia when absent is an inconvenience, but failing to detect it when present can be life-threatening. Because of this, artificial intelligence (AI) models used to assist such decisions are frequently trained with asymmetric loss functions that incorporate human decision-makers' trade-offs between false positives and false negatives. In two focal applications, we show that this standard alignment practice can backfire. In both cases, it would be better to train the machine learning model with a loss function that ignores the human’s objective and then adjust predictions ex post according to that objective. We rationalize this result using an economic model of incentive design with endogenous information acquisition. The key insight from our theoretical framework is that machine classifiers perform not one but two incentivized tasks: \emph{choosing} how to classify and \emph{learning} how to classify. We show that while the adjustments engineers use correctly incentivize choosing, they can simultaneously reduce the incentives to learn. Our formal treatment of the problem reveals that methods embraced for their intuitive appeal can in fact misalign human and machine objectives in predictable ways.

\end{abstract}

\newpage
\onehalfspacing
\setcounter{page}{1}   % Start page numbering at 1
\pagestyle{plain}      % Show page numbers in standard style
\section{Introduction}

As AI-based systems are increasingly deployed to make high-stakes decisions and execute consequential tasks autonomously, it is critical that they are `aligned' with the intentions of their human deployers  (\cite{bostrom2014align,bengio2023managing,ji2023ai, anwar2024tmlr}). If we take alignment at its simplest level to mean that systems are trying to behave as intended by their human deployers\footnote{This is the definition of \textit{intent alignment} (\cite{christiano2018clarifying}).}, alignment seems readily achievable in straightforward cases. Consider, for example, using a machine learner to detect potential cases of pneumonia from chest X-rays. While even the best human or machine classifier will misclassify some cases, the cost of errors is asymmetric: misdiagnosing pneumonia when it is absent (a false positive) inconveniences the patient; failing to diagnose pneumonia when it is present (a false negative) potentially threatens the patient's life. Accordingly, an aligned AI should behave like an ethical physician by erring on the side of overdiagnosing pneumonia. If medical ethics dictate that false negatives are 99 times as costly as false positives, it stands to reason that providing a machine learner with a loss function that penalizes false negatives 99 times as heavily as false positives should align the machine's actions with the deployer's intentions. 

Following this logic, machine learning models are frequently trained with asymmetric loss functions that codify experts’ assessed costs of false positives relative to false negatives.\footnote{For example, such asymmetric training for classifiers may be implemented through class weighting in the loss function or weighted resampling of the training data.} Implicit in these adjustments is what we term the \textit{aligned learning premise} (ALP): using the human's objective to train a machine learning model produces better performance in terms of that objective because it allows the human's objective to inform what the machine learns.

We show empirically that the ALP is false in two focal applications. In both cases, one would do better to first train the machine learning model using a standard loss function that ignores the human's objective and then adjust predictions ex post according to the human's objective, rather than to train with a \emph{utility-weighted} loss function that accounts for the human's objective, even though both loss functions are smooth and convex, which allows for optimization procedures to work effectively. In other words, trying to bake utility weights into training makes predictions worse --- even when judged by the utility-weighted objective itself.

\begin{figure}[ht]
    \centering
    \includegraphics[width=\linewidth]{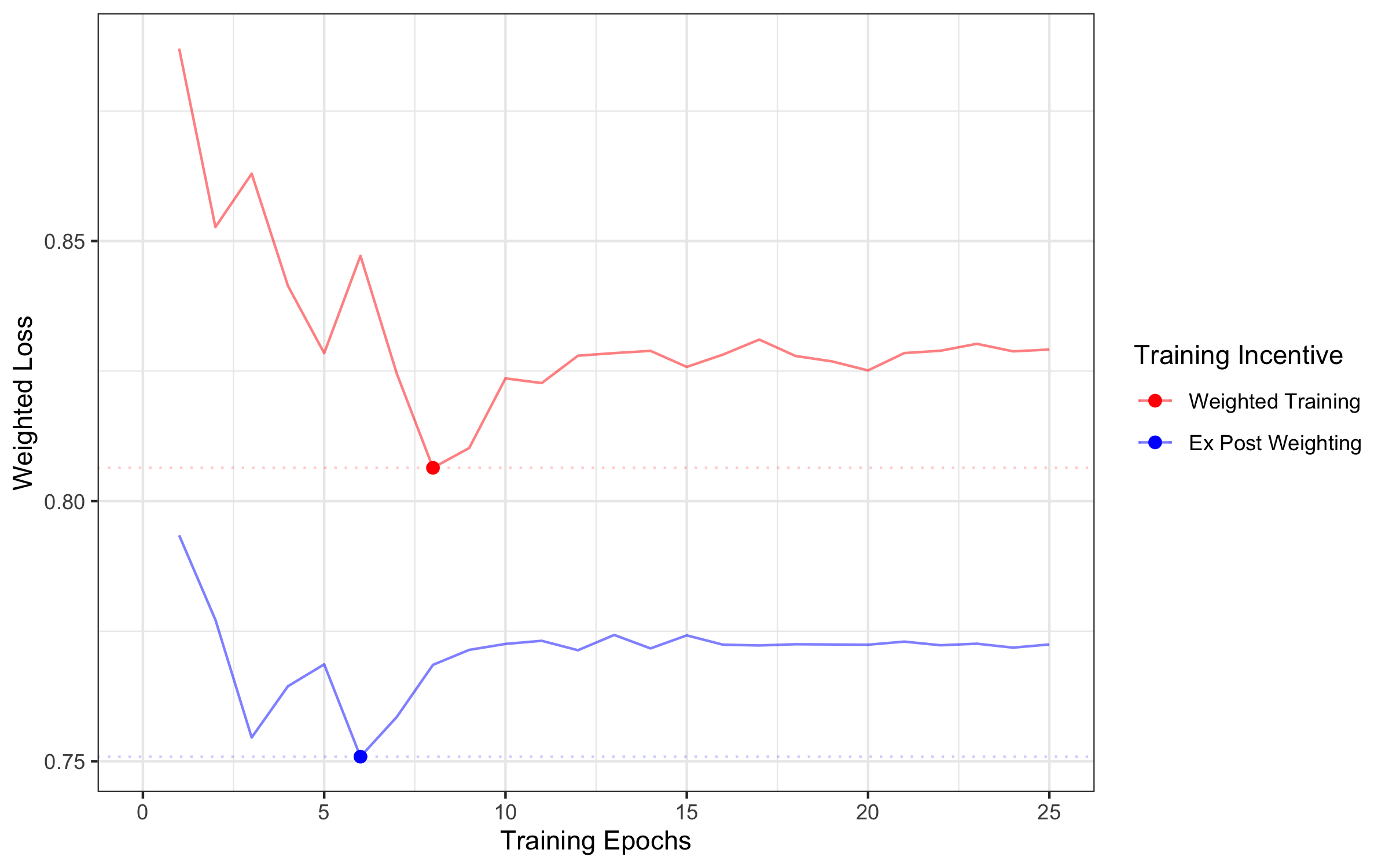}
    \caption{Weighted loss in the test sample across training incentives by training epoch, averaged across five runs. In red are the average weighted losses from training with the weights that reflect the human's objective, and in blue are the average weighted losses from training without weights, but accounting for the weights by transforming predictions ex post. Dots represent the minimum training epoch and dotted lines the corresponding weighted loss.}
    \label{fig:example1}
\end{figure}

These applications cover two standard prediction problems and algorithm architectures: medical diagnosis from chest X-rays using deep neural networks (\cite{rajpurkar2017chexnet}) and image classification in the CIFAR benchmark dataset using transformers (\cite{dosovitskiy2021image}). 
Previewing our results for pneumonia detection from chest X-rays, Figure \ref{fig:example1} plots the weighted loss for the same model under two possible loss functions: an unweighted one that does not account for the human's objective and one that weights instances of pneumonia relative to other classes at a ratio of 99 to 1.\footnote{As we discuss in Section \ref{sec:weight}, this weighting could be motivated by the importance of avoiding false negatives, issues of class imbalance in training data, or both.} In red are the weighted losses from training with the weights that reflect the human's objective (\textit{Weighted Training}).%
\footnote{The dark dots in each series correspond to the lowest utility-weighted loss across training epochs for the average run. In each case, the best results are obtained within the first ten epochs, suggesting the result is not attributable to underfitting and would be robust to variation in stopping rules. 
% In Section \ref{sec:emp} we disaggregate our results across multiple runs to illustrate that they are not explained by stochasticity in training.
} 
The ALP would suggest that this approach should produce the best performance one could achieve for the human's objective. However, in blue are the average weighted losses if we do not account for the human's objective when training the machine learning model, but instead account for utility by transforming predictions according to the desired objective (\textit{Ex Post Weighting}). Comparing these series shows that the machine minimizes utility-weighted losses not when trained with the utility-weighted loss function, but rather when trained without such weighting. In this pneumonia detection task, this dominance holds at every training epoch and across every run.

We rationalize these ALP failures using an economic model of incentive design with endogenous information acquisition. In our theoretical framework, alignment is achieved if the AI system makes predictions that maximize the human's expected utility. If the AI can learn perfectly (acquire a perfectly informative information structure), then the human only needs to adjust AI predictions ex post to achieve maximal expected utility. However, if the AI cannot learn perfectly, achieving alignment can depend on how the human designs the AI system.
% If more needed on this: \footnote{Possible reasons for partial learning include limited data, restrictions on the latent space, incorrect functional form assumptions, and optimization approximations such as early stopping.}
% Alternative wording: In our setting, it is also useful to observe that there is only scope for misalignment when learning is imperfect. In this case, different objectives can presumably lead to different prediction models and learning. 

One core design decision is which loss function to use when training the AI, and the ALP suggests that the optimal course of action is to base that decision on the human's own utility function. Viewing this as an incentive design problem, we ask if the human's utility function provides the correct incentives to the machine learner. The key insight provided by our model is that machine learners are performing not one but two incentivized tasks: choosing how to classify and learning how to classify.%
\footnote{Prominent classifier architectures that are separable in this way include regression trees (\cite{breiman1984class}), 
traditional neural networks (e.g., \cite{lecun1998nn}), and recent transformers (\cite{vaswani2017transformer}, \cite{dosovitskiy2021image})
} When the machine is \textit{choosing} how to classify a given X-ray, its loss function should guide it to output false positives 99 times as often as false negatives. Asymmetric weighting accomplishes this goal. But what incentives should the machine be given when \textit{learning} to classify X-rays? Intuition might suggest that learning is not an incentive problem: the machine should simply learn as effectively as possible. But the mathematics of machine learning dictate otherwise. Conventional machine learning algorithms learn to map features into classes through the process of gradient descent. Because the machine's loss function dictates the shape of the gradient, it necessarily shapes the machine's incentives for learning. The learning problem is therefore an incentive problem. 

Why does the human's objective not correctly incentivize the machine's learning problem? Formally speaking, why would the human's objective incentivize the machine to choose a poorly fitting information structure? We show theoretically that making a loss function asymmetric to account for the human's objective can backfire by weakening the machine learner’s payoff to substantive learning. Accounting for optimal ex-post adjustments in our theoretical and empirical results allows us to neutralize the impact of incentives for choosing and focus attention on the incentives for learning.

\begin{figure}[ht]
    \centering
    \includegraphics[width=\linewidth]{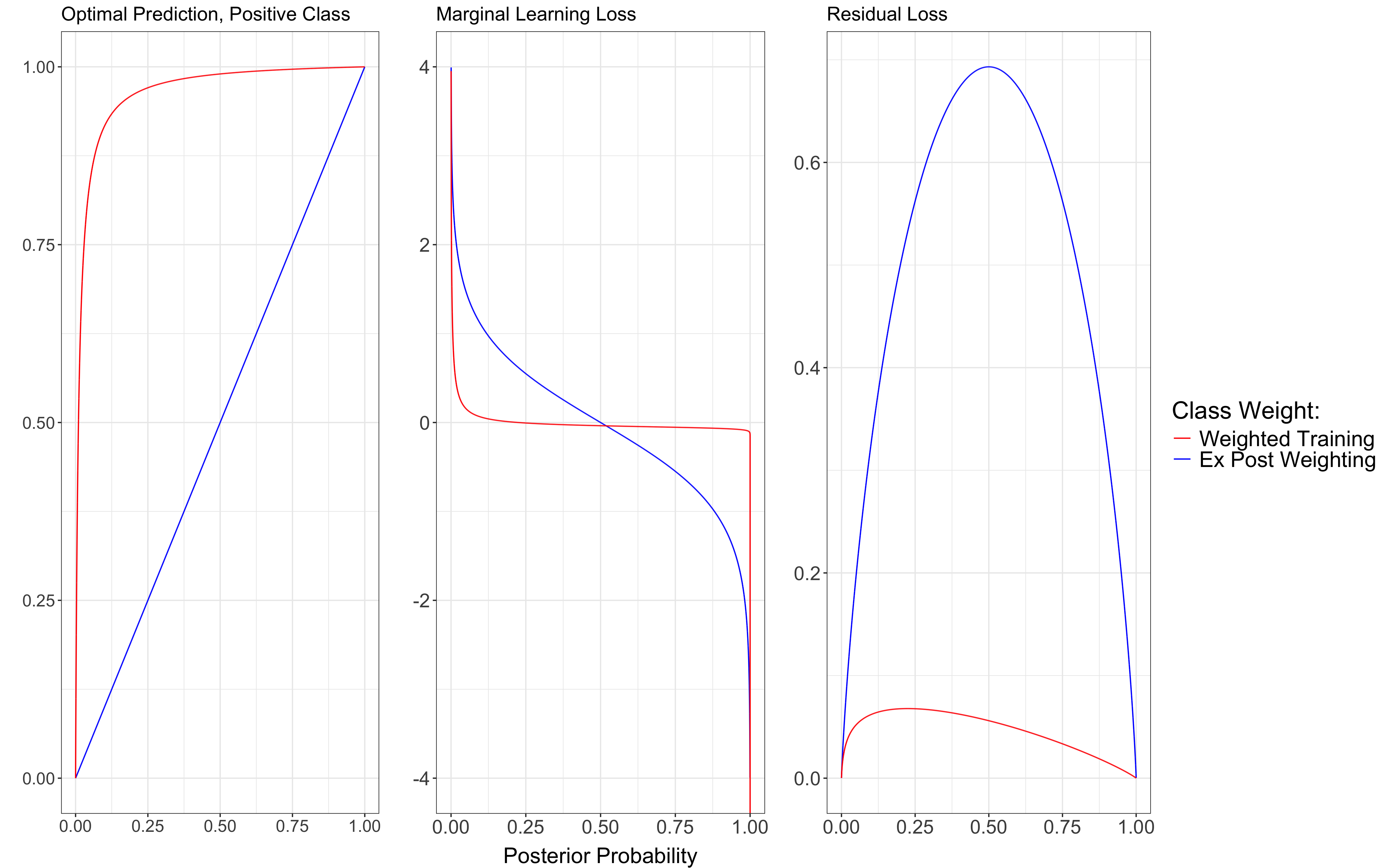}
    \caption{
        Incentives to choose and learn in the case of unweighted and weighted binary classification.
        Class weighting incentivizes distorting predictions as a function of posterior probabilities (left panel). 
        This distorts the marginal benefit of learning such probabilities (center panel), lowering and distorting the overall incentives for learning (right panel).}
    \label{fig:tri}
\end{figure}

Figure \ref{fig:tri} illustrates the forces behind our main theoretical results using the case of binary classification, such as classifying a patient with pneumonia. The left panel shows the machine's optimal prediction for each posterior probability of the ``positive'' class (e.g., pneumonia). Clearly, a machine that is incentivized to weight instances of pneumonia relative to other classes at a ratio of 99 to 1 would inflate predictions of pneumonia. The middle panel then shows how inflating predictions impacts the marginal incentives to learn: that is, the incentives to improve the posterior probability of the positive class. By inflating all predictions, the value of additional learning at all intermediate levels of learning is strongly dampened. Dampening the marginal incentives then lowers the overall incentive for learning---shown in the right panel---which is lower for all probabilities of the positive class. In summary, utility weighting provides correct incentives for choosing (left panel) but reduces the implied value of learning (middle and right panels), thus unintentionally misaligning human and machine objectives.

The widespread use of utility-weighted loss functions reflects both precedent and plausibility: they have delivered performance gains on benchmark datasets, and they mirror the trade-offs faced by decision-makers in domains like medicine or finance. Yet despite their popularity, these methods rest on an incomplete theoretical foundation. In particular, there has been no systematic treatment of how such practices interact with the incentives underlying machine learning, nor how they may alter the value of information acquisition. By providing an economic framework, we show that methods embraced for their intuitive appeal can in fact misalign human and machine objectives in predictable ways.

Our primary contribution is to illuminate the centrality of incentives in aligning machine behavior with human intentions. Using tools of economics, researchers have begun to consider how best to align machine outputs to human objectives, a process referred to as ``algorithmic design'' (\cite{liang2021algorithm}) and ``welfare-aware machine learning'' (\cite{rolf2020balancing}). \textcite{liang2021algorithm} consider the role of data inputs in achieving human fairness, and \textcite{rolf2020balancing} consider how to account for competing human objectives in machine learning. We add to this literature by showing how incentives play a central role in aligning machine outputs to human objectives. In this, we relate to previous work by
% Include only if we want econometrician referees: lieliwhite2010credit, elliottlieli2013predict, 
\textcite{hummelmcafee2017ads}, who consider the design of the training loss function based on downstream economic incentives.

Our paper also contributes to a literature that connects economic theory and machine learning (see \cite{liang2025usingmachinelearninggenerate} for a review). For instance, \textcite{samuelson2024constrained} and \textcite{aridor2020adaptive} use variational encoders to model how humans with cognitive constraints learn from the world. Our paper inverts this lens to investigate what models of human learning can tell us about how machines learn. That is, we explicitly model the machine's incentives and information acquisition as we would for a human. This approach links to a recent incentive design literature that considers how to best incentivize \emph{human} learners with positive learning costs (\cite{lambert2019elicitation,camara2025eliciting}).

By taking a Bayesian learning approach, our paper also connects machine learning with the information design literature (\cite{kamenica-gentzkow-2011bayesian, kamenica2019bayesian}). Specifically, we model the machine learner as first forming a probability distribution over classes and then selecting a prediction based on that distribution. We emphasize that this modeling approach is \textit{as if} because the machine learner does not necessarily follow such a process. Nevertheless, this abstraction allows us to connect what a machine learns with the classical definition and ordering of \textcite{blackwell1953equivalent}. 

Finally, the separation we make between AI predictions and final decisions appears often in the literature on fairness in AI (e.g., \cite{kleinberg2018discrimination}). While the problem studied in that literature is different (balancing outcomes across groups), our result is spiritually related to a robust finding in that literature, which is that it is better in terms of fairness concerns to train an unconstrained predictor and then post-process those decisions to make decisions that balance outcomes across protected categories (e.g., \cite{corbett2017algorithmic, menon2018cost, lipton2018does, kleinberg2018algorithmic, rambachan2020economic}). 

The rest of the paper proceeds as follows. Section \ref{sec:theory} provides our theoretical framework and results. Section \ref{sec:emp} provides the empirical results from our applications to chest X-ray diagnosis using deep neural networks and CIFAR image detection using transformers. 
Section \ref{sec:related} discusses how our methods and findings relate to machine learning literatures on cost-sensitive learning and alignment. Section \ref{sec:conclusion} concludes.

\section{Theoretical Framework and Results}\label{sec:theory}

This section formalizes the relationship between human objectives and machine learning. By framing the design of a machine’s loss function as an incentive design problem, we show why intuitive practices like utility-weighted loss functions can systematically backfire and, more fundamentally, uncover the two incentivized tasks of the machine: choosing how to classify a set of inputs given what it has learned; and learning from the inputs that it is given. While asymmetric weighting correctly shapes incentives for choice, we show that it can inadvertently weaken incentives for learning, leading to worse outcomes on the objectives that weighting was meant to serve. The framework that follows establishes this result formally and provides the foundation for the empirical applications that follow.

\subsection{Illustrative Example}\label{sec:example}

Imagine that a machine learning engineer (ME) needs to develop an \emph{AI system} to provide advice (an action $a$) based on an X-ray (a set of features $x$) that shows pneumonia or not (a class $y$). The ME wants the AI system to give ``good'' advice; that is, advice that is appropriate given the disease state. The ME's preferences for advice could reflect the preferences of the users of advice (doctors), the purchaser of the AI system (hospital), and/or the ME's employer (software company).

\begin{figure}
\centering
\resizebox{0.95\linewidth}{!}{%
\begin{tikzpicture}[
  >=Latex,
  node distance=16mm,
  box/.style     = {rounded corners, draw, very thick, align=center, inner sep=6pt, fill=none}, % <- fill=none to avoid covering children
  innerbox/.style= {rounded corners, draw, thick, align=center, inner sep=5pt, fill=white},
  lbl/.style     = {font=\footnotesize}
]

% Inner modules (placed first)
\node[innerbox, minimum width=34mm, minimum height=12mm] (f) {\textbf{AI model} \\ $f: \mathcal{X} \to \Delta(\mathcal{Y})$ };
\node[innerbox, right=16mm of f, minimum width=36mm, minimum height=12mm] (delta) {\textbf{Post processing} \\ $\delta: \Delta(\mathcal{Y}) \to \mathcal{A}$ };

% Outer program box (drawn after, but with fill=none)
\node[box, fit=(f)(delta), inner sep=8pt,
       label={[lbl]above:\textbf{AI system } $d: \mathcal{X} \to \mathcal{A}$}] (outer) {};

% IO anchors
\node[left=14mm of f] (Xin) {Features $x$};
\node[right=14mm of delta] (Aout) {Action $a$};

% Flow
\draw[->, very thick] (Xin) -- (f);
\draw[->, very thick] (f) -- node[above, lbl] {$\Delta(\mathcal{Y})$} (delta);
\draw[->, very thick] (delta) -- (Aout);

% % Composition note
% \node[below=6mm of outer.south, lbl] {$d \;=\; \delta \circ f$};

\end{tikzpicture}%
}
\label{fig:setup}
\caption{The components of the AI system.}
\end{figure}
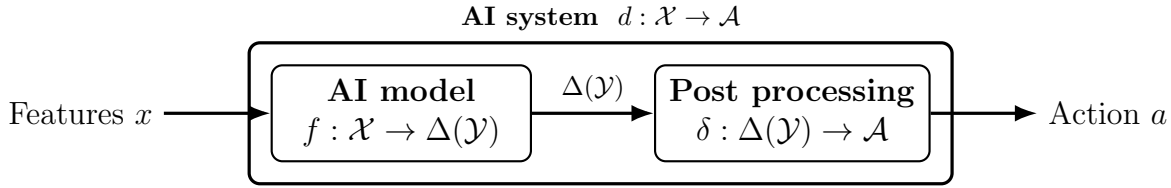

As illustrated in Figure \ref{fig:setup}, the AI system the ME develops can be summarized as a function $d$ from X-rays to advice. The ME trains an \emph{AI model} to be the key element in this AI system. The AI model outputs a probability of pneumonia based on an X-ray, and the AI model is summarized by a prediction function $f$. As a final step, the ME adds post processing $\delta$ to map the AI model output to specific advice $a$. 

In practice, the ME has to make many choices when training the AI model, but we focus on a particularly important one, which is the loss function $\ell$ used when training the AI model. In this setting, alignment is achieved if the ME's choice of the loss function $\ell$ produces an AI model $f$, which when combined with the optimal post processing $\delta$, generates an AI system $d$ that maximizes the ME's expected utility over advice and disease states given the set of possible X-rays.

\subsection{Human Decisions and Machine Predictions}

Formalizing this problem, human preferences are based on a finite set of actions $\mathcal{A} \equiv \{ 1, \dots, \na \}$ (e.g., types of advice) and a finite set of \emph{classes} $\mathcal{Y} \equiv \{1, \dots, \ny \}$ (e.g., one class might be pneumonia and another might be no disease). These preferences are summarized by a nonnegative utility function $\cl: \mathcal{A} \times \mathcal{Y} \to \mathbb{R}_+$. Our formal results require that the utility function $\cl$ is \textit{nondegenerate}, meaning that for every class there is some action that yields positive utility:%
\footnote{
This assumption is innocuous because affine transformations of the utility function preserve the same expected utility preferences, and so a utility function can always be transformed by addition of a positive constant to satisfy nondegeneracy. 
}%
\begin{equation}
\label{eq:nondegen-cost}
    \forall y \in \mathcal{Y}: \, 
    \exists a \in \mathcal{A}:
    \cl (a,y) > 0
\end{equation}

Human decisions are based on a set of features $\mathcal{X}$ (e.g., image pixels). The human decision-maker's problem is to develop an AI system $d: \mathcal{X} \to \mathcal{A}$ (shown in Figure~\ref{fig:setup}) to maximize expected utility $\mathbb{E} [ \cl (d(X), Y)]$.\footnote{Preferences can equally be summarized by a utility function to be maximized or a cost function to be minimized, and we consider the cost-based approach in Appendix \ref{apx:related} to facilitate comparison with the existing machine learning literature.} The expectation operator is defined with respect to a probability distribution $\mathbb{P}$, which is a random vector of features and classes $(X,Y)$ with realizations $(x,y)$ and finite support $\mathcal{X} \times \mathcal{Y}$. We interpret this probability distribution as an idealized infinite dataset with irreducible error.

To help develop this AI system, the human decision-maker employs machine learning to generate an AI model, which is a probabilistic prediction model. In the machine learning literature, this stage in developing an AI system is referred to as ``training.'' We focus on a particular aspect of training, which is the human decision-maker's choice of the machine's \textit{loss function} $\pl: \Delta (\mathcal{Y}) \times \mathcal{Y} \to \mathbb{R}$ over the set of classes $\mathcal{Y}$ and the set of probability distributions $\Delta (\mathcal{Y})$ over those classes. The human's ability to dictate the machine's loss function has motivated a variety of approaches in the machine learning literature on cost-sensitive learning and classification, which we discuss in Section \ref{sec:related} and detail in \autoref{apx:related}.

Our economic approach considers the human's choice of loss function as an incentive design problem for a downstream agent, the machine, which learns imperfectly. Based on the incentives provided by this loss function, the machine's problem is to find an AI model $f: \mathcal{X} \to \Delta 
(\mathcal{Y})$ to minimize expected loss
$\mathbb{E} [ \pl (f(X), Y)]$.

Finally, the human decision-maker combines the machine's AI model $f$ with post processing $\delta: \Delta (\mathcal{Y}) \to \mathcal{A}$ to create the AI system $d: \mathcal{X} \to \mathcal{A}$. In the machine learning literature, this stage in developing the AI system is referred to as ``inference.''
For an AI system that generates advice, post processing could involve simple recalibration of machine outputs (\cite{guo2017calibration,caplin2022calibrating}) or a calibrating coarsening of the machine outputs (\cite{hoong2025improving}). For an AI system that is a final decision-maker, post processing could be a threshold rule that produces a treatment decision. 

It is worth noting that the post-processing rule is only based on the probability distribution over classes. For many algorithms, this distribution is generated by applying the soft-max function to raw outputs. In principle, the designer could base the post-processing rule on other outputs from the machine, and it would be possible to extend our framework to allow for post processing to use an arbitrary space of latent representations.\footnote{We feel this is an interesting avenue for future work, and we thank Jakub Steiner for raising this point.}

\subsection{Alignment}

With this framework in place, we can formally define two forms of alignment. \textit{External alignment} is achieved if the loss function produces an AI model, which in combination with a decision rule, yields an AI system that maximizes the human's expected utility.
\textit{Internal alignment} is achieved if the loss function provided to the machine produces an AI model which maximizes that loss function. 

We focus our alignment framework on the common case of classification, such as diagnosing whether a patient has pneumonia. In this setting, the action is a class $\mathcal{A} = \mathcal{Y}$ and post processing is a classification rule $\delta: \Delta (\mathcal{Y}) \to \mathcal{Y}$. 
We focus on this case because it allows the connection between human preferences and machine learning to be stated as directly as possible, and it allows for simple expressions of theoretically optimal post processing. 
Nevertheless, our approach and results generalize.

\subsection{Optimal Post Processing}

We first use this framework to formally define optimal post processing for standard machine learning loss functions. Given a probability distribution $q \in \Delta (\mathcal{Y})$ over classes, the two key objects in this approach are the expected loss $\bar{\pl}$ of machine prediction $p \in \Delta (\mathcal{Y})$ and the expected utility $\bar{\cl}$ of human decision $y \in \mathcal{Y}$:
\begin{align}
\label{eq:expected-loss}
    \bar{\pl}(p,q) 
    &\equiv 
    \langle \pl (p, \cdot), q \rangle \\
\label{eq:conditional-risk}
    \bar{\cl} (y,q) 
    & \equiv 
    \langle \cl (y, \cdot), q \rangle
\end{align}
where $\langle \cdot, \cdot \rangle$ denotes the Euclidean inner product and $\pl (p, \cdot)$, $ \cl(a,\cdot)$ denote vectors over their second argument. For example, letting $p_y$ and $q_y$ denote the $y$-th elements of $p$ and $q$, $\pl (p,y) = - \log p_{y}$ is the logistic loss underlying cross-entropy loss $ \bar{\pl} (p,q) = - \sum_{y \in \mathcal{Y}} q_y \log p_y$.

For now, we consider a machine that has been given a loss function $\pl$
that is \textit{strictly proper} (\cite{buja2005loss, gneiting2007proper}), as with most standard loss functions (i.e., cross entropy and mean squared error).%
\footnote{
We do not consider the problem of how to select among proper scoring rules for prediction; for such results, see \textcite{buja2005loss} and also \textcite{hummelmcafee2017ads} for an economically motivated approach in the context of online advertising auctions.
}
For such a loss function, the optimal prediction is to output the true probability of each class:
\begin{equation}
    \label{eq:proper}
    q = \argmin_{p \in \Delta (\mathcal{Y})} \, \bar{\pl} (p,q) \quad \text{for all $q \in \Delta (\mathcal{Y})$}.
\end{equation}
As a result, optimal predictions will be \emph{calibrated}: when the model assigns a given probability to a class, that probability matches the observed frequency of the class.
When predictions are calibrated, the post processing that accounts for the human's preferences is the one that maximizes expected utility given the true probability of each class:
\begin{equation}
\label{eq:classification-conditional}
\delta^\cl (q) \equiv \argmax_{y \in \mathcal{Y}} \, \bar{\cl} (y,q)
\end{equation}

\subsection{Changing the Machine’s Loss Function}

Next we show how the machine's loss function can be designed to account for the human decision-maker's preferences. For this, we adopt an \textit{as-if} approach where we decompose the machine learner's prediction problem into two steps: (i) generating a probability distribution $q$, and (ii) converting those probabilities into optimal utility-weighted predictions $p^\cl (q)$. 
This approach is ``as if'' in the sense that the machine learner need not follow such a two-step procedure --- it only needs to generate predictions to optimize its objective.

We formalize \textit{utility-weighted loss} $\pl^\cl: \Delta (\mathcal{Y}) \times \mathcal{Y} \to \mathbb{R}$ as:
\begin{equation}
\label{eq:cost-loss}
\pl^\cl (p,y) \equiv 
\langle \pl (p, \cdot), \cl (\cdot, y) \rangle
\end{equation}
Conditional on a probability distribution $q$, the utility-weighted prediction $p^\cl$ minimizes expected utility-weighted loss: 
\begin{equation}
    \label{eq:prediction-conditional}
    p^\cl (q) \equiv \argmin_{p \in \Delta (\mathcal{Y})} \, \pl^\cl (p,q)
\end{equation}
Our main theoretical result establishes that utility-weighted loss provides the ``correct'' incentives for the machine when combined with optimal post processing, which is a simple argmax classification rule.

\begin{theorem}[Optimal Utility-Weighted Prediction]
\label{thm:alignment-outputting}
    Suppose that $\cl$ is nonnegative and nondegenerate and $\pl$ is a strictly proper loss function. For any class distribution $q \in \Delta (\mathcal{Y})$, the unique optimal utility-weighted prediction \eqref{eq:prediction-conditional} for each class $y$ is:
    \begin{equation}
    \label{eq:prediction-solution-conditional}
        p_{y}^\cl (q) = 
        \frac{
            \bar{\cl} (y,q)
        }{
            \rule{0pt}{2.2ex} \sum_{y' \in \mathcal{Y}} \bar{\cl} (y',q)
        }
    \end{equation}
\end{theorem}

\noindent
\autoref{thm:alignment-outputting} clarifies how to modify the machine's loss function to obtain a normalized expected utility \eqref{eq:conditional-risk} irrespective of the probability distribution $q \in \Delta (\mathcal{Y})$.
Consequently, preference-aligned decisions are obtained by combining a utility-weighted loss with a classification rule that selects the class with the highest normalized utility:
\begin{equation}
    \label{eq:dec}
    \delta (p^\cl) \equiv \argmax_{y \in \mathcal{Y}} \, p_y^\cl 
\end{equation}

\subsection{Learning Identity}
\label{apx:learning}

The preceding analysis establishes how human preferences can be embedded into the machine prediction problem via the utility-weighted loss function \eqref{eq:cost-loss} for utility-weighted predictions given a probability distribution over classes (Theorem \ref{thm:alignment-outputting}).
However, our analysis is thus far silent about what the machine is incentivized to learn.

We now consider how the utility-weighted loss function may \textit{implicitly} misalign incentives to learn. 
To do so, we adopt the classical approach of \textcite{blackwell1953equivalent} that operationalizes the value of learning about classes through the distribution over probability distributions over classes.
% In the following, we provide a simple yet general identity that decomposes any empirical loss in such terms. 

Treating predictor $f: \mathcal{X} \to \mathcal{Y}$ as fixed, let $P \equiv f(X)$ denote the random vector of resulting predictions and $(P,Y)$ the corresponding prediction evaluation data given predictor $f$ and raw data $(X,Y)$.
To make precise the problem decomposition and the meaning of ``what the machine learns,'' we introduce a simple yet useful identity following from the law of iterated expectations.

\begin{proposition}[Learning Identity]
\label{thm:decomposition}

Given a prediction loss function $\pl$ and evaluation data 
$(P,Y)$, the empirical risk can be expressed as:
\begin{equation}
\label{eq:decompose}
\mathbb{E} [ \pl (P, Y) ] = \mathbb{E} [ \bar{\pl} (P, Q)]
\end{equation}
where:
\begin{equation}
    \label{eq:what-is-learned}
    Q_y \equiv \mathbb{P} (Y = y | P) \quad \text{for all $y \in \mathcal{Y}$}
\end{equation}
is a random vector summarizing the information contained in the predictions $P$. % $h$ given data $(X,Y)$.

\end{proposition}

\noindent
% The random vector $Q^h$ defines a distribution over posterior class distributions which has rich precedent (e.g., \cite{blackwell1953equivalent}) as a summary of the information contained in hypothesis function $h$ given data $(X,Y)$; this is what we refer to as \textit{what is learned}.
The novel object of \autoref{thm:decomposition} is $Q$, which is a random vector of posterior class distributions that summarizes the information about classes inherent in the predictions $P$ given the actual outcomes $Y$ (on the same probability space).
To endow this information with meaning, recall that we implicitly assume the existence of some irreducible error.%
\footnote{
    In other words, we think of this as being evaluated in an uncountable test set, rather than a perfect fit in training with low external validity.
}
In that case we refer to $Q$ as a summary of \textit{what is learned}; such a distribution of posterior probability distributions is a well-known classical means of conveying information content, with close ties to information value (\cite{blackwell1953equivalent}).

Thus, \autoref{thm:decomposition} decomposes the machine's prediction objective (i.e., finding a predictor function $f$ to minimize empirical risk) into an ``as if'' sequence of problems. 
The learning problem is to find an informative $Q$.
Then, as a function of any realization $q$ of what is probabilistically learned $Q$, the conditional prediction problem is to choose a prediction $p(q)$ to minimize the expected prediction loss $\bar{\pl} (p, q)$.
% ; the optimal such cost-sensitive predictions were derived in \autoref{thm:alignment-outputting}.
Again, note that for calibrated predictions, what is learned is immediately given by the predictions themselves, $Q = P$.
However, the value of expression \eqref{eq:what-is-learned} is that $Q$ has meaning even when the predictor is systematically miscalibrated and $Q \neq P$, as arises by design in \autoref{thm:alignment-outputting} when prediction losses are weighted by preferences. 

This approach allows us to model and understand the (dis)incentives for learning through the (marginal) incentives for learning different distributions $q$ over the class space.%
\footnote{
Another potential comparative static is changes in distributions over probability distributions rather than changes in the probabilities themselves.
However, this higher-order object also entails a much richer set of directional derivatives, and it would still be composed of these simpler constituent parts.
}

\subsection{Implicit Learning Incentives} 
\label{sec:learn}

To understand how learning incentives are affected by human preferences $\cl$, we define the \textit{(utility-weighted) residual learning loss} as the expected utility-weighted loss of probability distribution $q$, once it is optimally post processed according to \eqref{eq:prediction-solution-conditional}:
\begin{equation}
\label{eq:indirect-learning}
\tilde{\pl}^\cl (q) \equiv 
% \sum_{y \in \mathcal{Y}} \pl^\cl (p^\cl (q), y) \times q_y 
\bar{\pl}^\cl (p^\cl (q), q)
\end{equation}
The next result derives a simple expression for the \textit{marginal} residual learning loss at probability distribution $q$ as the vector of utility-weighted losses at the optimal prediction across classes.

\begin{proposition}[Incentives Underlying Learning]
\label{thm:incentives-learning}
As in Theorem \ref{thm:alignment-outputting}, suppose that $\cl$ is nonnegative and nondegenerate and $\pl$ is a strictly proper loss function.
Then residual learning loss $\tilde{\pl}^\cl$ is a concave and differentiable function, and its gradient at each $q$ is the vector of utility-weighted prediction losses evaluated at the optimal utility-weighted prediction. Expressed component-wise,
\begin{align}
\frac{\partial \tilde{\ell}^u}{\partial q_y} (q) = \ell^\cl (p^\cl (q), y)
% \nabla_q \, \tilde{\pl}^\cl (q) &= \pl^\cl (p^\cl (q), \cdot)
\label{eq:indirect-gradient}
\end{align}
\end{proposition}

\noindent

\subsection{Special Case: Class-Weighted Cross Entropy}
\label{sec:cw-ce}

The common case of class-weighted cross entropy is a specialization in two ways. 
First, human preferences are utility-weighted indicators for correct classification:
\begin{equation}
\label{eq:class-utility}
    u^w(y',y) \equiv w_y 1 \{ y' = y \}
\end{equation}
with $w \in \mathbb{R}_{++}^\ny$ ensuring nondegeneracy \eqref{eq:nondegen-cost}. 
This yields \textit{class-weighted loss} $\pl^w (p,y) \equiv w_y \pl(p,y)$,
which is a special case of utility-weighted loss.
Second, the base loss function is taken to be the standard one for classification: logistic loss, $\pl(p,y) = - \log p_y$.
Combining these specializations, weighted loss \eqref{eq:cost-loss} becomes:
\begin{equation}
\label{eq:cw-cross-entropy}
\pl^w(p,y) = - w_y \log p_y % \quad \text{for all $p \in \Delta (\mathcal{Y})$}
\end{equation}
The optimal class-weighted prediction \eqref{eq:prediction-solution-conditional} of class $y$ given class distribution $q$ becomes:
\begin{equation}
    \label{eq:optimal-prediction-w}
    p_{y}^w (q) = \frac{ 
        w_{y} q_{y}
    }{
        \langle w, q \rangle
        % \sum_{y' \in \mathcal{Y}} (w_{y'} q_{y'})
    }
\end{equation} 
The gradient of learning loss \eqref{eq:indirect-gradient} at probability distribution $q$ becomes:
\begin{equation}
    \label{eq:gradient-w}
    \frac{\partial \tilde{\pl}^w }{\partial q_y} (q)
    % = \pl^w (p^w (q), y) 
    = - w_y \log p_y^w (q)
\end{equation}

The gradient of learning losses \eqref{eq:gradient-w} captures the direct and indirect effects of weighting on learning, assuming that the machine makes optimal predictions given what it has learned. 
Directly, a higher weight $w_y$ on a class $y$ increases the value of shifting probability distributions (i.e., learning) on that class. 
Indirectly, however, a higher weight $w_y$ decreases the remaining term $- \log p_y^w (q)$ as the prediction $p_y^w (q)$ increases toward 1.
Intuitively, as a class $y$ becomes more heavily weighted, the corresponding prediction $p^w (q)$ becomes more detached from the probability distribution $q$. Namely, it becomes relatively more beneficial to make a high prediction on class $y$ regardless of underlying information. 
In the limit case where $w$ is only nonzero in its $y$-th element, it is optimal to output $p_y^w (q) = 1$ regardless of probability distribution $q$, so that learning is not valuable at all. 

To develop further intuition for how the incentives for learning are affected by the choice of weights $w$, consider the binary-class case ($m = 2$).
In that case, the probability distribution $q$, weights $w$, and optimal predictions $p^w (q)$ can be parameterized respectively by the scalar probability $q_1$, the scalar weight $w_1 = 1-w_2$, and the scalar prediction $p_1^{w_1} (q_1)$ on class 1 (the positive class), with the analogs on class 2 being the complementary probabilities.%
\footnote{
The class weight normalization $w_1 + w_2 = 1$ in this theoretical subsection is scaled down by a factor of two relative to our normalization \eqref{eq:cost-normal} in the experimental section \ref{sec:emp}. 
In the former case, intuition is simplified with complementary probabilities; in the latter case, rescaling relative to unweighted cross entropy (where weights sum to the number of classes $m$) disentangles the effect of reweighting on the shape of the objective function from effects on its magnitude.
}
Then we can express the marginal learning loss $\tilde{\pl}^{w_1} (q_1)$ as: 
\begin{equation}
\label{eq:deriv-w}
    \frac{d \tilde{\pl}^{w_1} }{dq_1}  (q_1) = - w_1 \log (p_1^{w_1} (q_1)) + (1- w_1) \log (1 - p_1^{w_1} (q_1))
\end{equation}

Returning to Figure \ref{fig:tri} from the Introduction, we plot the optimal class-weighted prediction $p_1^{w_1} (q_1)$, the marginal learning loss $\frac{d}{dq_1} \tilde{\pl}^{w_1} (q_1)$, and the normalized learning loss $\tilde{\pl}^{w_1} (q_1)$ as a function of probability $q_1$ for two class weights: the ``Ex Post Weighting'' case $w_1 = w_2$ and a ``Weighted Training'' case that emphasizes the positive class $w_1 = 99 w_2$. We call the former ``Ex Post Weighting'' because optimal post processing accounts for the human's utility weights.

The left-hand subfigure shows that emphasizing the positive class extremizes and thus flattens the optimal prediction as a function of the probability on much of the domain; in particular, it becomes optimal to output a high prediction on the weighted class except in cases of very low probabilities. 
The middle subfigure shows that when learning about the class has less bearing on the predictions, the marginal value of learning is decreased and even negligible on much of the domain. 
The right-hand subfigure shows that this flattens the learning loss.

This intuition extends beyond the binary-class case. 
The derivative \eqref{eq:deriv-w} is a difference in the partial derivatives in the gradient \eqref{eq:gradient-w}, and this derivative is small because its constituent terms are small. 
More generally, suppose class $1$ is (heavily) over-weighted. 
Then the partial derivative of $\tilde{\pl}^w (q)$ with respect to $q_1$ is low on most of the support because $p_1^w (q)$ is close to 1 regardless of the probability $q$, so that $\log p_1^w (q)$ is close to 0 and the marginal value of learning for losses is still small even though the weight $w_1$ is large; however, the partial derivatives on other classes are also low because the weights $w_{-1}$ are small.  
In the extreme case where the loss function puts all weight on the predictions of class 1, there is no incentive to learn at all.

It is worth noting that while the incentive for learning appears suppressed for weighted learning in Figure \ref{fig:tri} relative to unweighted learning, it turns out that marginal learning loss in Figure \ref{fig:tri} is \textit{higher} for extreme probabilities. Thus, it is conceivable that for settings where initial learning is very easy, weighted learning would provide optimal learning incentives. 

It is also worth noting that it is not possible to overcome the dampened incentives for learning by uniformly increasing the size of the weights. In practice, a constant rescaling of the machine loss function would be offset by a corresponding change in the learning rate to avoid issues around numerical precision. Unlike human learners, the machine simply follows a path of stochastic gradient descent over a high dimensional surface toward a locally optimal solution. Thus, we interpret our analysis as illustrating distortions in this learning surface, which we would only expect to increase in more complex multi-class settings.

\subsection{Establishing Misalignment}

Let $P$ and $P^\cl$ denote the random vectors of predictions obtained through training with unweighted and utility-weighted loss functions, respectively. We now formalize both external and internal alignment within our framework.

Then we would conclude that incorporating human preference misaligns the machine output (according to internal alignment) if the optimal predictions from unweighted training outperform the optimal predictions from weighted training in terms of weighted losses:
\begin{equation}
\label{eq:misalign}
    \mathbb{E} [ \pl^\cl (p^\cl (P), Y) ] < \mathbb{E} [ \pl^\cl (P^\cl, Y)]
\end{equation}
Thus, we measure the strength of internal alignment as the gain (reduction) in weighted loss from using unweighted loss in training. Note that this test does not modify the utility-weighted predictions; rather it constructs a potential improvement from the unweighted predictions by transforming those according to the optimal utility-weighted prediction \eqref{eq:prediction-solution-conditional}. 

Analogously, we would conclude that incorporating human preferences misaligned the machine output (according to external alignment) if the corrected predictions from unweighted training outperform the weighted predictions in terms of human utility (here ``classification utility''):
\begin{equation}
\label{eq:misalign-external}
    \mathbb{E} [ \cl (\delta(p^\cl(P)), Y) ] > \mathbb{E} [ \cl (\delta(P^\cl), Y) ]
\end{equation}
Following the above, we measure the strength of external alignment as the gain (increase) in expected utility from using unweighted loss in training.

Next, we establish the presence of such misalignment in a variety of applications and show that it has consequences for the downstream classification objective. 

\section{Experiments}
\label{sec:emp}

We document machine misalignment with human objectives by revisiting two prominent multi-class classification applications and architectures: 
chest X-ray diagnosis with deep neural networks (\cite{rajpurkar2017chexnet}, \cite{wang2017chestxray}) and CIFAR image classification with vision transformers (\cite{dosovitskiy2021image}, \cite{krizhevsky2009cifar}). 
In each application, we focus on human utility functions \eqref{eq:class-utility} corresponding to class-weighted loss emphasizing one class $y$ with a ratio $w_y = 99 w_{-y}$ relative to any other class $-y$.
In Appendix \ref{apx:related}, we explain how this can be motivated by either an enhanced importance of avoiding false negatives on that class (e.g., failing to detect pneumonia) or as a way of addressing class imbalance (e.g., pneumonia cases constitute approximately 1 percent of the training data in the application of \cite{rajpurkar2018deep}).\footnote{
In the imbalanced case of chest X-rays, we also consider training according to inverse probability-weighted cross entropy $w_y\mathbb{P}(y) = w_{y'} \mathbb{P}(y')$ for any classes $y,y'$, where the probability weights are computed in the training data to avoid referencing test or validation data.}
Across our applications, we normalize the weight vector $w$ so that the expected loss of agnostically outputting the prior probability $\mu \in \Delta (\mathcal{Y})$ is fixed across training schemes and relative to unweighted loss:%
\footnote{
    In the balanced CIFAR data, such weights sum to the number of classes $m$; that is, their average value is one. 
    The average value is not as easily interpretable in the imbalanced chest X-ray data. 
}
\begin{equation}
    \label{eq:cost-normal}
    \bar{\pl}^w (\mu,\mu) = \bar{\pl} (\mu, \mu)
\end{equation}
This normalization is intended to hold fixed the initial expected loss --- before the machine learns to discern images ---
and thus to ensure that our results are not driven by systematic differences in the magnitude of the gradient across weighting schemes or hyperparameters such as the learning rate. 
% avoids conflating effects due to the shape of the learning objective with effects due to the magnitude of the objective, for example through a misspecified learning rate.
% Besides our manipulation of the loss function, we largely follow standard parameter values and protocols as detailed in Appendix \ref{sec:data}.
Besides our manipulation of the loss function, we follow standard parameter values and protocols. 
Each model is trained using an NVIDIA Tesla A100 80GB GPU on the Google Colab platform. 
We discuss further details and preexisting implementations in Appendix \ref{sec:data}.

We compare two training regimes: first, training according to class-weighted cross entropy \eqref{eq:cw-cross-entropy} [\textit{Weighted Training}], and second, training according to unweighted cross entropy and adjusting the predictions ex post according to \eqref{eq:optimal-prediction-w} [\textit{Ex Post Weighting}].
We compare these predictions according to the machine's own class-weighted loss in a test dataset to evaluate objective misalignment \eqref{eq:misalign}. 
In addition, we compare performance according to the human's classification utility \eqref{eq:class-utility}. 
However, we would expect smaller differences in classification utility across regimes since categorical decisions will vary across training regimes less often than the underlying probabilistic predictions.

We also compare the training regimes across a variety of emphasized classes. 
In the chest X-ray task that uses the data from \textcite{wang2017chestxray}, we separately study four classes with varying levels of occurrence in our training data (Pneumonia = 0.003, Cardiomegaly = 0.012, Pneumothorax = 0.024, and Infiltration = 0.105).  
In the image classification task using CIFAR data (\cite{krizhevsky2009cifar}), we focus on emphasizing the most difficult class (cat images in CIFAR-10 and maple tree images in CIFAR-100) since the baseline predictor and classifier achieve near-perfect performance already on other classes, leaving little scope for potential performance improvements through aligned learning. 
In each variant, we repeat the training procedure five times to separate substantive (mis)alignment from the inherent stochasticity of the training procedures.
In the chest X-ray tasks, we evaluate the trained models in the test data at the end of every training epoch, and in the CIFAR applications, we evaluate the trained model in the test data at every 100 training steps. 
Henceforth, we refer to the training epochs and training steps as \emph{training intervals}.

\subsection{Results}
\begin{table}[!htbp]
\centering
\renewcommand{\arraystretch}{1.2}
\begin{tabular}{@{} ll c c c @{\hspace{10pt}} c c c @{}}
\toprule
\multicolumn{2}{c}{} &
\multicolumn{3}{c}{\textbf{Weighted Loss}} &
\multicolumn{3}{c}{\textbf{Classification Utility}} \\
\cmidrule(r){3-5} \cmidrule(l){6-8}
\multicolumn{2}{c}{} &
% \textit{\small Weighted Training} & \textit{\small Ex Post Weighting} &
% \textit{\small Weighted Training} & \textit{\small Ex Post Weighting} \\

\textit{Weighted} & \textit{Ex Post} & \textit{\%} & 
\textit{Weighted} & \textit{Ex Post} & \textit{\%} \\
& &\textit{Training} & \textit{Weighting} & \textit{Gain} & 
\textit{Training} & \textit{Weighting} & \textit{Gain} \\

\midrule

\multicolumn{8}{l}{\textbf{Weight: Pneumonia (Chest X-ray)}} \\
& Mean & 0.805 & 0.749 & 6.96 & 0.292 & 0.309 & 5.82 \\
& Min  & 0.796 & 0.743 & 6.66 & 0.284 & 0.296 & 4.23 \\
& Max  & 0.812 & 0.755 & 7.02 & 0.296 & 0.318 & 7.43 \\
\addlinespace[0.75ex]

\multicolumn{8}{l}{\textbf{Weight: Cardiomegaly (Chest X-ray)}} \\
& Mean & 0.401 & 0.383 & 4.49 & 0.333 & 0.333 & 0.00 \\
& Min  & 0.398 & 0.377 & 5.28 & 0.332 & 0.328 &-1.20 \\
& Max  & 0.408 & 0.391 & 4.17 & 0.337 & 0.336 &-0.30  \\
\addlinespace[0.75ex]

\multicolumn{8}{l}{\textbf{Weight: Infiltration (Chest X-ray)}} \\
& Mean & 0.214 & 0.211 & 1.40 & 0.597 & 0.597 & 0.00 \\
& Min  & 0.212 & 0.210 & 0.94 & 0.597 & 0.596 &-0.17 \\
& Max  & 0.215 & 0.213 & 0.93 & 0.597 & 0.597 & 0.00 \\
\addlinespace[0.75ex]

\multicolumn{8}{l}{\textbf{Weight: Pneumothorax (Chest X-ray)}} \\
& Mean & 0.370 & 0.347 & 6.22 & 0.355 & 0.363 & 2.25 \\
& Min  & 0.360 & 0.343 & 4.72 & 0.353 & 0.360 & 1.98 \\
& Max  & 0.377 & 0.352 & 6.63 & 0.361 & 0.366 & 1.39 \\
\addlinespace[0.75ex]

\multicolumn{8}{l}{\textbf{Weight: Inverse Probability (Chest X-Ray)}} \\
& Mean & 0.723 & 0.696 & 3.73 & 0.125 & 0.130 & 4.00 \\
& Min  & 0.719 & 0.688 & 4.31 & 0.123 & 0.129 & 4.88 \\
& Max  & 0.729 & 0.703 & 3.57 & 0.127 & 0.131 & 3.15 \\
\addlinespace[0.75ex]

\multicolumn{8}{l}{\textbf{Weight: Cat (CIFAR-10)}} \\
& Mean & 0.033 & 0.024 & 27.3 & 0.988 & 0.989 & 0.10 \\
& Min  & 0.032 & 0.023 & 28.1 & 0.986 & 0.989 & 0.30 \\
& Max  & 0.035 & 0.025 & 28.6 & 0.989 & 0.990 & 0.10 \\
\addlinespace[0.75ex]

\multicolumn{8}{l}{\textbf{Weight: Maple Tree (CIFAR-100)}} \\
& Mean & 0.188 & 0.159 & 15.4 & 0.960 & 0.969 & 0.94 \\
& Min  & 0.181 & 0.156 & 13.8 & 0.955 & 0.968 & 1.36 \\
& Max  & 0.195 & 0.161 & 17.4 & 0.964 & 0.970 & 0.62 \\
\addlinespace[0.75ex]

\bottomrule
\end{tabular}
\caption{Comparison of prediction and evaluation methods under different weighting schemes. Each row reports mean, min, and max losses across five runs.
The column ``\% Gain'' reports the percentage improvement of Ex Post Weighting relative to Weighted Training for each row.
The gain is positive for improvement: lower weighted loss or higher classification utility. 
}
\label{tab:main}
\end{table}

Table \ref{tab:main} summarizes the performance results across training procedures and applications in terms of weighted loss (used to measure internal misalignment) and classification utility (used to measure external misalignment).
For each application, the table presents basic summary statistics (mean/min/max) of the optimal performance (over training intervals) in the test sample across five training runs.
In every application, we find that the \textit{Weighted Training} procedure is outperformed on average by the \textit{Ex Post Weighting} procedure of training without weights. 
Indeed, in nearly every application, the worst performance on the weighted training objective of \textit{Ex Post Weighting} exceeds the best performance of \textit{Weighted Training}. The one exception is Infiltration diagnoses in chest X-rays, which is also the case where the mean performance improvements of \textit{Ex Post Weighting} are smallest. 

Consider our leading application of pneumonia: across five runs, we consistently achieve better performance on the weighted training objective under the \textit{Ex Post Weighting} procedure than under \textit{Weighted Training}, with a mean reduction in loss of 6.96\%.%
\footnote{Computed as $((0.749 - 0.805) / 0.805) \times 100 \%$ in the first panel of Table \ref{tab:main}.} Importantly for our alignment interpretation, our measure of outperformance is benchmarked to the \textit{Weighted Training} objective.

These results also largely translate to external alignment, which relates to the human's utility (the downstream objective of maximizing the underlying classification utility), although as expected the differences are smaller and noisier given the discontinuous nature of the objective. 
Classification performance according to the \textit{Ex Post Weighting} procedure performs either better or approximately the same on average (up to three decimal points, in the cases of Cardiomegaly and Infiltration diagnoses). 
In Appendix \ref{sec:figs} we illustrate this classification performance improvement across training intervals in pneumonia diagnosis. 

\subsection{Discussion}

In summary, we document consistent performance improvements on both the machine's own objective and the human's downstream classification utility when the machine is \textit{not} trained according to the weighted objective but instead when its unweighted predictions are corrected ex post to accord with the performance objective. 
In Section \ref{sec:theory}, we presented a theoretical framework and arguments attributing this objective misalignment to implicit incentives that distort or even stifle the value of learning substantive information. 

Our main findings are consistent with prior Gauss-Markov style results showing that, in the absence of heteroskedasticity, an unweighted objective and estimator are better than weighted alternatives (\cite{greene2012econometric}). 
This alternative explanation has some important shortcomings, however. Most critically, it would predict that unweighted learning should outperform weighted learning without ex post corrections. In Appendix \ref{sec:figs} we also compare weighted loss and classification utility from \textit{Weighted Training} to an unweighted training regime without modifying predictions (\textit{Unweighted Raw}), and we find that \textit{Weighted Training} significantly outperforms \textit{Unweighted Raw} according to the weighted objectives.

\section{Related Machine Learning Literature}
\label{sec:related}
Our paper seeks to bridge a longstanding literature on cost-sensitive learning and class imbalance with an active literature on the alignment of machine learning to human preferences. 
Our overarching objective is to illuminate the centrality of \textit{incentives} in aligning machine actions with human intentions. By embedding choice incentives into learning, many common methods conflate the objectives of choosing and learning, thereby distorting one while aligning the other. 

Our incentive-based approach is based on a novel cost-weighted prediction loss function and generalizes and unifies many of the existing solutions in the literature, including (partial or binary-class) solutions based on thresholding, reweighting, resampling, and base rate adjustments. 
Additionally, our incentive-based approach yields a general multi-class formula for analytical recalibration of cost-weighted predictions. 

\subsection{Cost-Sensitive Learning}
Motivated by the suboptimal performance of standard classifiers in cases of cost-sensitive classification and class imbalance, the literature on cost-sensitive learning built on a series of workshops at the turn of the century, including a workshop on cost-sensitive learning at the 2000 International Conference on Machine Learning (ICML, \cite{dietterich2000cost}), a workshop on learning from imbalanced data at the 2000 Association for the Advancement of Artificial Intelligence (AAAI) meetings (\cite{japkowicz2000imbalance}, \cite{provost2000imbalance}, \cite{japkowiczholte2001workshop}), and a second workshop on learning from imbalanced data at the 2003 ICML (\cite{chawla2003imbalance, drummondholte2003imbalance, maloof2003imbalance, chawlajapkowiczkolcz2003imbalance}).
\cite{fernandez2018learningimbalanced} provides a recent review. 
The challenge of cost-sensitive classification and its resolution are embodied in \cite{elkan2001cost}: ``the essence of cost-sensitive decision-making is that it can be optimal to act as if one class is true even when some other class is more probable.''
% To the best of our knowledge, the fundamental questions of generally separating and relating prediction and classification incentives, as well as how to incorporate cost sensitivity into prediction in a simple, unified, yet multi-class way, have remained open.

% There are two potential aspects of this problem which it is important to distinguish (e.g., \cite{provost2000imbalance}). 
% The first is that an algorithm that maximizes accuracy or minimizes the error rate by 
% applying zero-one loss to probabilistic outputs may optimally never correctly label unlikely classes, even though in many applications these are exactly the classes for which correct labeling is most important (e.g., positively identifying a cancer or fraud).

Our multi-class incentive-based method grounded in \eqref{eq:cost-loss} and the theory of proper scoring%
\footnote{
Our approach departs from this literature by aggregating proper scoring evaluations according to misclassification costs to generate \textit{im}proper scoring rules. 
In contrast, \cite{buja2005loss} \textit{decomposes} proper scoring rules as weighted sums of cost-weighted misclassification errors to select among \textit{proper} scoring rules. Their approach is in turn based on the proper scoring decompositions of \textcite{shuford1966admissible, savage1971elicit, schervish1989prob}.
}
% We refer to our approach as incentive-based classification. 
generalizes and clarifies existing approaches in the literature, including reweighting training data (e.g., \cite{breiman1984class, domingos1999metacost, drummondholte2000costinsensitive, elkan2001cost, ting2002cost, zhouliu2010weight}), resampling training data (e.g., \cite{kubat1997imbalance, elkan2001cost, drummondholte2003imbalance, zadrozny2003cost, azy2004multi, xia2009sample}), and ex post prediction adjustments based on modified base rates (\cite{elkan2001cost, saerens2002adjust}) or thresholding and conditional risk minimization (\cite{domingos1999metacost, elkan2001cost, margineantu2002costclass}).%
\footnote{
Similar strategies for designing cost-sensitive binary classifiers have also been considered in the econometric literature, e.g., 
\textcite{lieliwhite2010credit, elliottlieli2013predict} and references therein.
}
% Ex post rescaling: For regression trees, \cite{drummondholte2000costinsensitive} find that combining standard training (leaf splitting criterion) with cost-sensitive ex post operation (pruning and leaf-labeling) performs as well as other methods. 
Our simple framework combines the generality of existing data-based resampling solutions with the efficiency of algorithm-based solutions. It also generalizes existing analytical recalibration approaches beyond the well-studied binary resampling case (\cite{pozzolo2015calibration}), and it makes precise the connections between cost-sensitive learning (i.e., incentives) and class imbalance (i.e., the distribution of classes and features in training data) in a general multi-class framework. 
We detail these relations in Appendix \ref{apx:related}.

Empirically, the literature has debated the utility of introducing costs into the machine's loss functions.
For example, \textcite{elkan2001cost} conjectures that introducing costs into the training objective may not significantly affect performance relative to introducing costs into the classification incentive, and \textcite{vanderschueren2022} conclude that it is more important that costs are included in the decision-making strategy than whether they are included in the training or classification stage. 
Our work is, to the best of our knowledge, the first to demonstrate that introducing such costs into the training objective may be counterproductive with respect to the terminal goal because it is counterproductive for the machine's own learning objective.
Our theoretical framework and results provide the foundation to develop improvements by framing and decomposing the problem as one of incentive design (compared to, e.g., resampling strategies).

\subsection{Alignment}
As machine learning models have become more powerful and foundational, researchers have raised substantial concerns about the risks and possibilities of misalignment between human values and machine objectives (e.g., \cite{bostrom2014align, russell2016align, amodei2016safety}). As famously presaged by \textcite{wiener1960align}: ``If we use, to achieve our purposes, a mechanical agency with whose operation
we cannot interfere effectively\dots we had better be quite sure that the purpose put into the machine is the purpose which
we really desire.''
Seemingly the central challenge of alignment, then, is correctly and precisely communicating complex objectives about which we ourselves may not yet be sure. 
However, while aligned objectives are almost certainly necessary for human-machine alignment, another question is whether they are sufficient.
Our results suggest that even a ``correctly'' specified human objective may encode other, possibly pathological, implicit incentives into machine learning.

The goal of our method is to align the machine's prediction incentives with the human decision-maker's classification incentives. 
Our as-if modeling approach generates a natural separation into an \textit{inner} and \textit{outer} alignment problem, analogous to \textcite{hubinger2019mesa}. 
In the inner alignment problem, the model 
% is incentivized to optimize predictions as a function of the prediction-embedded misclassification costs and the latent posterior class distribution.
is incentivized to transform posterior class distributions into predictions that align with misclassification costs.
In the outer alignment problem, the machine is incentivized to learn a mapping from features to (posterior distributions over) classes, given the transformation inherent in the inner alignment problem.  

Our empirical and theoretical finding on calibration and learning incentives can be rephrased using this inner/outer alignment dichotomy: 
Given outer alignment, cost-weighting may still affect the shape of the inner objective in a way that softens incentives toward inner alignment, even though it does not distort those incentives.
% even when the inner objective (cost-sensitive output) is aligned, it may induce pathological incentives in the outer objective (actual machine learning),
Conversely, alignment of the inner objective (cost-sensitive outputs) may induce pathological incentives in the outer objective (actual machine learning),
such that the machine incentives that theoretically align with human objectives are \textit{not} optimal for training the machine.
In our case, the paradox is that even the machine would prefer --- according to its objective function --- to be trained according to other incentives, conditional on adjusting its outputs to reflect its incentives to choose.
While this may be rationalized as a consequence of modifying the shape of the loss function in counterproductive ways, we reiterate that our loss function transformations at least preserve the smoothness and convexity properties that motivate surrogate loss.
Furthermore, we would expect the possibility of inadvertent incentive spillover to become more severe in settings that are more complex or where even the inner objective is misaligned.
Thus, we caution that not all human objectives may be productively encoded into the machine learning algorithm, even when they can be precisely articulated.
% \textit{even if} outer incentives are correct! (Korinek and Balwit 2022).

% At the risk of interpretational overreach, we find it intriguing to further expand on the analogies and connections between our as-if modeling approach and current approaches to studying alignment.
Analogously to \textcite{hubinger2019mesa}, our approach to the inner problem studies the model as if it were an optimizer, given posterior probabilities over classes; we do not, however, claim that the model in the inner problem is a \textit{mesa} optimizer, which seems reserved for reinforcement learning, large language, or general foundation models with richer input and output spaces introducing the possibility of in-context learning and optimization. 
% Not to quote ultimately but useful food for thought from Olsson, connecting mesa optimization and in-context learning ``In the longer run, concepts such as mesa-optimization or inner-alignment postulate that meaningful learning or optimization could occur at test time (without changing the weights). In-context learning would be an obvious future mechanism for such hidden optimization to occur, whether or not it does so today. Thus, studying in-context learning seems valuable for the future.''
% in-context learning: I.e., Is a capability learned (in context), or already latent and just elicited? 
Still, cost-sensitive classification is perhaps a minimal extension of standard classification for which the task is complex enough for the two-stage analogy to be non-tautological,%
\footnote{
Tautological in their sense that any object's objective can always be defined as being and behaving like what it is.
% In the non-cost sensitive case, the optimization is just truthful (and thus trivial) revelation of beliefs, which is similar to Hubinger's example of a bottle cap's tautological behavioral objective of optimizing the objective of behaving like a bottle cap.
}
yet simple enough where we can articulate the correct incentives for the machine, as well as the procedure for recovering the machine's behavioral objective or implicit preferences (analogous to inverse reinforcement learning; \cite{ngrussell2000irl}).
% Leike et al's ``reward-result'' gap: the difference between the ``reward model'', i.e., base objective, and the ``reward function recovered with perfect inverse reinforcement learning.'' 
Since we analytically derive the inner as-if objective, we can thus conclude that 
% Then Hubinger's \textit{inner alignment problem} is the problem of eliminating the base-mesa objective gap, and outer alignment is the problem of eliminating the gap between the base objective and the intended goal of the programmers. 
% ``In the context of machine learning, outer alignment refers to aligning the specified loss function with the intended goal, whereas inner alignment refers to aligning the mesa-objective of a mesa-optimizer with the specified loss function.''
% Distinguish the mesa-objective from the behavioral objective, which is the objective that appears to be optimized by the system's behavior, which can be operationalized as the objective recovered via perfect inverse reinforcement learning.
% We can recover a meaningful objective by estimating the best fit of loss terms, also related to inverse calibration...
% Also though, ``the fact that a system's behavior results in some objective being maximized does not make the system an optimizer.'' It is at least somewhat less obvious in the cost-sensitive case than just standard classification..% 
% cost-sensitive learning creates perhaps the simplest split between beliefs and actions.
% Cost-sensitive classification a special domain where we can make precise the \textit{correct} incentives for the machine.
% \textit{as if} an inner layer of optimization as a function of beliefs.
% Reframing the objective as a function of an emergent attribute, namely implicit posteriors or what is learned, rather than emergence of an objective function per se
the inner objective is essentially (albeit more softly) aligned upon using our approach, and show why this still expectedly creates challenges for aligning the outer objective. 
We are also able to use our analytical approach to define and recover latent but important attributes such as ``what the machine has learned.'' 
This suggests the potential for as-if modeling approaches to complement deepened mechanistic intuition (e.g., \cite{olsson2022induction, vonoswald2023mech, vonoswald2024mech}) and pre-formal analysis (e.g., \cite{ngo2024align}) for understanding, interpreting, and aligning ever-more-complex machine learning models with human preferences.

While our results are broadly consistent with the empirical findings of \textcite{vanderschueren2022} and \textcite{caplin2022calibrating} in the case of binary classifiers, we document strict and significant suboptimality in the multi-class case, suggesting that implicit misalignment increases with the complexity of the prediction problem. 

\section{Conclusion}
\label{sec:conclusion}

This paper examines a simple premise that underpins much current practice: if human decision-makers value some errors more than others, then training machine learners on a utility-weighted loss should better align model behavior with human objectives. We show empirically that this aligned learning premise (ALP) can fail in systematic ways. We explain this failure theoretically using economic principles of incentive design. Machine learners simultaneously face two incentivized tasks: \emph{choosing} how to classify given what they know and \emph{learning} what is worth knowing in the first place. While utility weighting can correctly incentivize the choice, it inadvertently weakens incentives for learning by flattening the value of additional information.

Our theoretical framework formalizes this separation. Modeling prediction as a two-step process, forming posterior probabilities and then mapping them to predictions, we derive the optimal way to embed human preferences into the prediction step (Theorem~\ref{thm:alignment-outputting}). We then show how this transformation changes the shape of the learning objective. Weighting inflates preferred classes, which then reduces the marginal value of moving posteriors in most regions of the state space (Proposition~\ref{thm:incentives-learning}). In short, the same adjustment that aligns incentives for what to choose can misalign incentives for what to learn.

Across two standard applications --- chest X-ray diagnosis with deep neural networks and CIFAR image classification with transformers --- we find consistent evidence in favor of a simple alternative: train with a strictly proper, unweighted loss to learn calibrated probabilities, and then impose human objectives ex post based on those objectives. This \textit{Ex Post Weighting} approach dominates training directly on the utility-weighted objective (\textit{Weighted Training}) when evaluated on that very objective, and it typically yields equal or better downstream classification utility.

Our analysis focuses on multi-class classification under proper losses and nondegenerate human utilities. It abstracts from dynamic, interactive, or sequential settings, and from additional issues such as fairness or robustness to distribution shift. Extending incentive design to these domains (e.g., by constructing training objectives that both preserve information incentives and respect complex human goals) is an important direction for future work.

The high-level message from this paper is that alignment should not focus exclusively on determining the human's objective; it must also provide the machine learner with the right incentives to acquire information. 
Recognizing both alignment objectives may prove crucial to building AI systems that are not, in effect, misaligned by design.

\newpage
\printbibliography

\newpage

\appendix

\section{Relations to Cost-Sensitive and Imbalanced Learning}
\label{apx:related}

% {\color{red} Now that we've moved this into an appendix, use cost rather than utility only in this subsection for ease of reference with the existing literature.}

In this section, we use our general incentive-based approach to compare, contrast, and unify four classes of solution to cost-sensitive learning and class imbalance: (i) thresholding, (ii) reweighting, (iii) resampling, and (iv) base rate adjustments. We then use our  approach to (iv) derive a novel and general multi-class formula for analytical recalibration across these contexts. 

Contrasting with Section \ref{sec:theory}, we mainly represent downstream preferences and objectives as nonnegative and nondegenerate costs $c$ rather than utilities $u$ to facilitate comparison with the literature.  
It is straightforward to show that our results and approach hold upon replacing expected utility maximization with conditional cost (i.e., risk) minimization. 
% It is straightforward to show that the theoretical alignment result (Theorem \ref{thm:alignment-outputting}) holds whether we model human preferences via costs to be minimized or utilities to be maximized.

\subsection{Thresholding and Conditional Risk Minimization}

Thresholding refers to the practice of changing the decision cutoffs for classification, and conditional risk minimization to its appropriate multi-class generalization. 
In its simplest form, this means to train a prediction model without reference to cost, 
% so that predictions are probabilistic $p (q) = q$, 
and then classify instances based not on their immediate predictions, but on their cost-derived conditional risk. 

In the binary-class case $\mathcal{Y} = \{ 0,1 \}$ where class distributions $q$ are summarized by the positive-class probability $q_1$, \textcite{elkan2001cost} derives a closed-form solution for the positive-class probability $q_1^c \in [0,1]$ where the conditional risks are equal, $\bar{c} (1,q^c) = \bar{c} (0,q^c)$.
% \footnote{
%     Equation 2.
% }
This probability $q_1^c$ in turn serves as a threshold for classifying positive instances: 
\[
y^c (q_1) \stackrel{(1)}{=} 1 \{ q_1 \geq q_1^c \} \stackrel{ (2)}{=} 1 \{ \bar{c} (1, q_1) \leq \bar{c} (0,q_1) \}.
\]
In the multi-class case, there is no simple threshold determining classification as in the first equality, yet the conditional risk minimization inherent in the second equality still applies and is generalized by the conditional risk minimizer (in our leading case of utility, the expected utility maximizer \eqref{eq:classification-conditional}).
This conditional risk minimizer formula appears previously in \textcite{margineantu2002costclass}, who also notes that such ex post conditional risk minimization allows general application of a single trained model across misclassification cost functions.
The Metacost algorithm of \textcite{domingos1999metacost} applies the same conditional risk minimization formula \eqref{eq:classification-conditional} in an interim training stage in order to reclassify the model \textit{training} data, before training the final algorithm on the reclassified data.
Thus, Metacost departs from unweighted learning by applying conditional risk minimization at the training rather than at the classification stage, which requires retraining an algorithm for any change in misclassification costs.

\subsection{Reweighting}
\label{sec:weight}

Reweighting training instances changes their relative importance in the loss function.
% emphasis
% Reweighting an existing distribution of training labels to achieve an objective.
By implicitly changing training incentives,
%in the algorithm, 
reweighting is spiritually similar to our incentive-based approach; nevertheless, it is only a special case
confined to having a single weight per instance. 
Consider, for example, the common case of \textit{class weighting} discussed in Section \ref{sec:cw-ce}.
The corresponding \textit{class-weighted prediction loss} is:
\begin{equation}
\label{eq:Loss-class-weight}
    \pl^w (p,y) \equiv w_y \pl(p,y)
\end{equation}
which is evidently a special case of our preference-weighted loss.
Again, the optimal class-weighted prediction of class $y$ given class distribution $q$ becomes:%
\begin{equation}
    p_{y}^w (q) = \frac{ 
        w_{y} q_{y}
    }{
        \langle w, q \rangle
        % \sum_{y' \in \mathcal{Y}} (w_{y'} q_{y'})
    }
\end{equation} 
We now show how our novel approach relates to and extends existing class-reweighting approaches.
We discuss the connections to problems of class (im)balance in \autoref{sec:imbalance}.

In the binary classification case, Theorem 1 of \textcite{elkan2001cost} considers how to make a given target probability threshold on the positive class $p_1^*$ correspond to a given probability threshold $p_1^o$ for an accuracy-maximizing classifier. 
In our notation, Elkan's problem is to choose weights on the negative class $w = (w_0,1)$ such that:
\[
    q_1 \geq p_1^* \iff p_1^w (q_1) \geq p_1^o
\]
By monotonicity of $p_1^w (q_1)$ in its argument $q_1$, this is equivalent to solving: 
\[
    p_1^w (p_1^*) = p_1^o
\]
which yields Elkan's formula: 
\[
    w_0 = \frac{p^*}{1-p^*} \frac{1-p_0}{p_0}
\]
The threshold approach is not generalizable to the multi-class case. 
In contrast, our approach of beginning with (mis)classification preferences $c$ or $\cl$ and incorporating them into training incentives via the preference-weighted prediction loss is generally applicable, with even multi-class weighting \eqref{eq:Loss-class-weight} as a special case.

In the multi-class case with general misclassification costs $c$, \textcite{breiman1984class, domingos1999metacost, ting2002cost, drummondholte2003imbalance} propose reducing the cost matrix $c(y',y)$ to a cost vector $c(y) = \sum_{y'} c(y',y)$ that can be used for class weighting.
The resulting machine incentives are significantly different from those of our incentive-based approach because:
\[
    \pl (p,y) c(y) = \pl(p,y) \sum_{y'} c(y',y) \neq \sum_{y'} \pl(p,y') c(y',y)  \equiv \pl^c (p,y)
\]
Indeed, in the same multi-class case with general misclassification costs $c$, \textcite{zhouliu2010weight} propose an alternative class-weighting scheme $w$ to address the issues of the preceding cost-reduction proposal, which instead solves:
\[
    \frac{w_y}{w_{y'}} = \frac{c (y',y)}{c (y,y')} \quad \text{for all $y,y' \in \mathcal{Y}$.}
\]
If such a weighting scheme does not exist, they decompose the multi-class problem into a set of binary-class problems and apply the preceding constraints separately. 
Even if such a weighting scheme does exist, it typically disagrees with our approach since the ratio of weights between classes is determined only by a pair of misclassification costs, rather than a pair of conditional risks which may depend non-trivially on the entire vector of losses given a true label.
In contrast to their approach, our solution (i) generally exists, (ii) yields closed-form prediction formulas, and (iii) is rigorously grounded in prediction and classification incentives via \autoref{thm:alignment-outputting}.

More generally, it is evident from comparing the set of cost-weighted and class-weighted predictions \eqref{eq:optimal-prediction-w} that our incentive-based approach is not recovered by class weighting alone.
Indeed, our approach naturally generalizes the existing practice of weighting observations only by their true class. 
For example, evaluating weighted loss for the logistic loss function yields a new family of weighted loss functions we term \textit{doubly-weighted cross entropy}:%
\footnote{
This weighting procedure is equally applicable to other proper loss functions, such as the Brier score (\cite{brier1950score}). 
}
\begin{equation}
    \label{eq:doubly-weighted}
    \pl^c (p,y) = 
    - 
    \langle \log p, c(\cdot, y) \rangle
\end{equation}
This allows for differentially costly misclassifications across labels, given a true class $y$. 
For example, some medical misdiagnoses may be less costly if they lead to similar courses of action, or image misclassifications may be less costly if the mislabels are still similar in some characteristic space.

Finally, another noteworthy but not directly comparable family of reweighting procedures for cost-sensitive learning involve boosting, where instance weights are chosen dynamically during training \parencite{fan1999costboost, sun2006boostingimbalance}.

\subsection{Resampling}
\label{sec:sample}

Resampling is intuitively similar to the preceding case of reweighting, except that it manipulates the training data rather than the training incentives (i.e., the algorithm). 
In the binary classification case, the problem is well-studied (e.g., \cite{breiman1984class, kubat1997imbalance, japkowicz2000imbalance, elkan2001cost}), particularly in the context of class imbalance to which we return in \autoref{sec:imbalance}.
% Also called up-sampling and down-sampling --- in binary context, respectively replicating cases from the minority or ignoring cases from the majority. 

Existing resampling approaches in the multi-class case reveal a particular advantage of resampling relative to reweighting: a single training instance can be repeatedly resampled, whereas it can be reweighted only once (unless it is also resampled).
Most closely related to our approach, \textcite{xia2009sample} provide a theoretically grounded data expansion technique based on \textcite{azy2004multi}, in which a single training instance may be repeatedly resampled in proportion to its various cost entries.
More specifically, given a distribution of features and classes $\mathbb{P} (x,y)$, they derive a resampled distribution $\hat{\mathbb{P}} (x,y)$ such that the 
minimizer of cost-sensitive classification on distribution $\mathbb{P}$ is theoretically equivalent to the minimizer of zero-one classification on distribution $\hat{\mathbb{P}}$.
Thus, their solution based on manipulating training data is similar to our solution based on manipulating incentives. 
Still, our incentive-based approach confers several relative advantages. 

As noted previously, an advantage of resampling relative to reweighting is the possibility of repeatedly resampling the same training instance, which allows resampling to convey more complex incentives than those permitted by reweighting. 
At the same time, a disadvantage of resampling is its inefficiency from both a data and computational standpoint, due to data loss from not sampling all observations, alongside redundancy, stochasticity, and increased complexity from sampling observations repeatedly. 
Our incentive-based approach captures the relative advantages of both reweighting and resampling --- allowing for complex (mis)classification preferences without any training data manipulation.

In addition, our incentive-based approach yields a simple implementation and intuition via the cost-weighted prediction loss and closed-form solutions for predictions, which can also be used for analytical recalibration or belief recovery (\autoref{sec:calibration}) as well as understanding the resulting implicit incentives to learn (\autoref{sec:learn}). 
Additionally, our approach is simply extended to \textit{example-based costs} (\cite{zadrozny2003cost}, \cite{azy2004multi}) by allowing the cost function in our incentive-based approach to also depend on features, $c(x,y,y')$, in which case the conditional risk and the adjustment formula will also depend on features $x$.
On the other hand, our approach is inherently based in proper surrogate loss functions, which may limit its use to certain algorithm classes; in contrast, resampling can be generally implemented, even if its theoretical justifications are equally limited. 

\subsection{Base Rate Adjustments and Class Imbalance} 
\label{sec:imbalance}

The resampling and reweighting approaches discussed in the preceding \autoref{sec:weight} and \autoref{sec:sample} are often motivated by questions of class imbalance (for overviews, see \cite{chawlajapkowiczkolcz2004imbalance} and \cite{fernandez2018learningimbalanced}). 
% \cite{kubat1997imbalance, japkowicz2000imbalance, drummondholte2003imbalance, chawla2005imbalance, chawla2004imbalance})
In turn, class imbalance and its resolution relate closely to the idea of modifying class base rates.
For example, in settings of binary classification, a common intuition is that standard algorithms are biased toward the majority class, which can be addressed by either upweighting or upsampling the majority class, and/or downweighting or downsampling the minority class.
We have already resolved the question of correcting asymmetric human classification incentives in the machine prediction problem (\autoref{thm:alignment-outputting}), and our incentive-based approach was notably independent of the actual class distribution. 

Still, the incentive-based approach is also productive for the goal of effectively changing base rates: namely, given overall class base rates $\bar{q} \in \Delta (\mathcal{Y}) \cap \mathbb{R}_{++}^m$ in training, how do we modify prediction incentives such that the machine is incentivized to choose \textit{as if} the base rates were instead $\bar{p} \in \Delta (\mathcal{Y})$?
\textcite{saerens2002adjust} provide an adjustment formula for posterior probabilities when only base rates are modified, which is rooted in Bayes' rule.
We recall this in our framework and language below.%
\footnote{
In the binary case, a similar formula appears in Theorem 2 of \textcite{elkan2001cost}.
The formula also appears previously in the economic literature on Bayesian persuasion (\cite{alonsocamara2016het}).
}

\begin{proposition}[\textcite{saerens2002adjust}]
\label{thm:imbalance}
    For a fixed set of conditional feature probabilities $\mathbb{P} (X = x | Y = y)$, the conditional class distributions $ p(x), q(x) \in \Delta (\mathcal{Y})$ of feature vector $x$ under respective target base rates $\bar{p} \in \Delta (\mathcal{Y})$ and source base rates $\bar{q} \in \Delta (\mathcal{Y}) \cap \mathbb{R}_{++}^m$ are related by formula:
    \begin{equation}
        \label{eq:class-balance}
        p_y (x) = 
        \frac{ 
            \bar{w}_{y} q_{y} (x)
        }{
            \langle \bar{w}, q (x) \rangle
            % \sum_{y' \in \mathcal{Y}} \bar{w}_{y'} \times q_{y'} (x)
        }
        \text{ for all $y \in \mathcal{Y}$},
    \end{equation}
    where $\bar{w}_y = \bar{p}_y / \bar{q}_y$.
\end{proposition}

\noindent
It is immediate by comparison of \eqref{eq:class-balance} to class-weighted adjustments \eqref{eq:optimal-prediction-w} and the underlying class-weighted prediction loss \eqref{eq:Loss-class-weight} that the base rate adjustment formula is implementable by our incentive-based approach with class weights $\bar{w}_y = k \bar{p}_y / \bar{q}_y$ for any positive constant $k$.
% Alternatively, the analytical correction can be applied ex post to the set of posteriors $q(x)$ across features $x$ generated by an algorithm under base rates $\bar{q}$.

% A naive intuition is the following: in the absence of any learning and given a proper (unweighted) loss function $L$, it is optimal to output distribution $\bar{q}$ to obtain expected payoff $\sum_{y \in \mathcal{Y}} [\bar{q}_y \times L(\bar{q},y) ]$; similarly, if the class distribution were $\bar{p}$, it would be optimal to output distribution $\bar{p}$ to obtain expected payoff $\sum_{y \in \mathcal{Y}} [\bar{p}_y \times L(\bar{p},y) ]$. 
% Comparison to \eqref{eq:Loss-class-weight} of these expected payoffs and their implicit weights  suggests then the use of class weights equal to the ratio of source and target priors, $w_y = \bar{p}_y / \bar{q}_y$.
% However, this intuition is incomplete without a further model of how predicted class distributions are generated and dependent on base rates, especially since the loss function $L(q,y)$ is typically non-linear in its prediction argument $q$.
% It may not even be obvious ex ante that a simple incentive-based solution to rebalancing base rates generally exists. 
% The following result clarifies and confirms the sense in which it does.

A special case is the practice of inverse class weighting $\bar{w}_y = 1 / \bar{q}_y$, which normalizes the base rates to be uniform across classes.
Whether this provides the correct classification incentives depends on the classification preferences and the distribution of test data.
% unless the classification incentive coincides with choosing the class that would be most likely if base rates were balanced.
Indeed, our incentive-based approach provides a useful means of simultaneously unifying and disentangling the related problems of cost-sensitive misclassification and base rate adjustment strategies to address class imbalance.
% formalize the sense in which class imbalance is a special case of cost-sensitive learning. 
% framework seems helpful for resolving/unifying class imbalance (distribution of data) and cost sensitivity (differing costs).

\subsection{Inversion and Calibration}
\label{sec:calibration}

% : Probability Recovery and Calibration

When predictions are probabilistically correct, they are said to be \textit{calibrated}.
Probabilistic model calibration has clear advantages, as evidenced by the large literature devoted to its study (e.g., \cite{platt1999calibration, zadroznyelkan2001, zadroznyelkan2002, guo2017calibration, minderer2021calibration}). 
By design, however, the predictions generated by incentivizing the machine according to (mis)classification preferences are typically \textit{mis}calibrated because preference-weighted predictions distort latent probabilities:
\[
p^\cl (q) \neq q
\]
This raises a question about when it is possible to analytically recover calibrated probability estimates from observed, cost-weighted predictions. 
\autoref{thm:recalibrate} provides a simple yet general analytical solution when the utility function $u: \mathcal{Y} \times \mathcal{Y} \to \mathbb{R}_+$ can be represented as an invertible matrix $U$ with entries $U_{y',y} = u(y',y)$. 
In this case, let $U^{-1}$ denote the inverse matrix of $U$, and let $\bar{\cl}^{-1}(y,p)$ denote the expected utility of $y$ given prediction $p$ and (inverse) utility function $\cl^{-1}$ defined by matrix $U^{-1}$. Then:

\begin{theorem}[Analytical Recalibration]
\label{thm:recalibrate}
    For a nonnegative, nondegenerate classification utility $u$ represented as an invertible matrix $U$ and a strictly proper loss function $\ell$, the optimal cost-weighted prediction function \eqref{eq:prediction-solution-conditional} is invertible with closed-form:
    \begin{equation}
        \label{eq:calibration}
            [p_y^{\cl}]^{-1} (p) = \frac{
                \bar{\cl}^{-1} (y, p)
            }{
                 \rule{0pt}{2.2ex} \sum_{y' \in \mathcal{Y}} \bar{\cl}^{-1} (y', p)
            }
    \end{equation}
    where $u^{-1}$ is the utility function defined by inverse matrix $U^{-1}$.
\end{theorem}

\noindent
\autoref{thm:recalibrate} provides a simple means to recover latent probabilities $q$ underlying observed predictions $p^\cl (q)$ as long as the utility matrix is invertible: compute the inverse utility matrix, and then compute the normalized risks of the observed predictions according to the inverse matrix.  
In the special case of class weighting with a diagonal matrix defined by weights $w \in \mathbb{R}_{++}^m$, the matrix inverse $U^{-1}$ is just the inverse of every diagonal entry, and so \eqref{eq:calibration} reduces to:
\begin{equation}
    \label{eq:invert-cw}
    [p_{y}^w]^{-1} (p) = \frac{ 
        (p_{y} / w_{y})
    }{
        \sum_{y' \in \mathcal{Y}} (p_{y'} / w_{y'})
    }
\end{equation}
Even this special multi-class case generalizes existing binary analytical calibration methods such as \textcite{pozzolo2015calibration}.
% , caplin2022calibration

When costs are not invertible, multiple conditional probability distributions can generate the same prediction, resulting in a non-recoverable loss of probabilistic information. 
Still, given the observed prediction data $(P \equiv f(X),Y)$, it remains possible to empirically recalibrate predictions by fitting a model for the empirical class probabilities:
\[
\hat{q}_y (p) \equiv \mathbb{P} (Y = y | P = p),
\]
e.g., \textcite{platt1999calibration, zadroznyelkan2001, zadroznyelkan2002, guo2017calibration}). 
Even when underlying probabilities are analytically recoverable from observed predictions, whether the resulting probabilities are indeed well-calibrated is an application- and algorithm-specific empirical question. 

A related but subtly different question is whether predictions are well-calibrated to their incentives, i.e., \textit{loss-calibrated} (\cite{caplin2022modeling}).
When the cost matrix is invertible, this can be checked by seeing whether the analytically recalibrated predictions \eqref{eq:calibration} are empirically well-calibrated in the traditional sense, according to standard methods (e.g., \cite{degroot1983comparison, niculescu2005predicting, naeini2015ece, nixon2019ace}).  
A more general approach that does not require analytical invertibility of the adjustment formula is the following. 
The predictions are \textit{loss-calibrated} if the empirical predictions $P$ coincide with the analytical prediction $p^\cl (\cdot)$ evaluated at the empirically recalibrated beliefs $\hat{q} (P)$:
\begin{equation}
\label{eq:loss-calibration}
    p^\cl (\hat{q} (P)) = P
\end{equation}
% This recovers the standard multi-class definition of calibration (e.g., \cite{minderer2021calibration}) in the cost-insensitive case $p^C (q) = q$: 
% \[
%     \mathbb{P} (Y = y | P ) = P_y
% \]
In the case of a proper loss function \eqref{eq:proper} where there are no incentives to misreport, i.e., $p^\cl (q) = q$ for all $q \in \Delta (\mathcal{Y})$, loss calibration collapses to standard multi-class calibration (e.g., \cite{minderer2021calibration}):
\[
    \hat{q} (P) = P
\]
Definition \eqref{eq:loss-calibration} of loss calibration does not rely on invertibility of the adjustment formula $p^\cl (\cdot)$, only that there exists an accurate empirical procedure or model for recovering recalibrated beliefs $\hat{q} (p)$. 
In particular, loss calibration permits multiple latent ``subjective'' posterior probabilities to map to the same observed prediction.
Since the observed $\hat{q} (p)$ will be a convex combination of such latent posteriors at every realization $P = p$, validity of this definition requires that level sets of the adjustment formula $p^\cl (\cdot) $ be convex.
We confirm this in the following lemma. 
\begin{lemma}
\label{thm:convex-closed}
    For any nondegenerate and nonnegative classification utility $\cl$, level sets of the analytical adjustment formula \eqref{eq:prediction-solution-conditional} are convex. That is, for any $q,q' \in \Delta (\mathcal{Y})$ and $\alpha \in [0,1]$, 
    \[
        p^\cl (q) = p^\cl (q') = p \implies p^\cl (\alpha q + (1-\alpha) q') = p
    \]
\end{lemma}

\noindent
Thus, \eqref{eq:loss-calibration} is a valid definition of loss calibration, in the sense that if it is violated, there exists some latent subjective posterior $q$ for which the observed prediction is not well-calibrated to its incentives. 
Loss calibration is testable similarly to calibration, since $p^\cl (\cdot)$ is a known, deterministic function that can be applied ex post to the recovered probabilities $P$.

Finally, we note that determining whether a prediction model is loss-calibrated is essentially the same as determining whether its actual objective --- as summarized by a utility function $u$ --- coincides with its behavioral objective --- as summarized by a utility function $\hat{u}$ \textit{estimated} from the prediction data $(P,Y)$, e.g., by multivariate \textcite{platt1999calibration} scaling, and analogously to the inverse reinforcement learning exercise (\cite{ngrussell2000irl}). 
In turn, this is analogous to establishing a zero reward-result gap (\cite{hubinger2019mesa, leike2018align}) in the machine's prediction problem.

% \subsection{Advantages}
% Advantages of approach:
% First, it provides a simple approach to alignment. 
% Second, an analytical correction
% Third, allows us to recover a latent set of such beliefs
% Fourth allows us to decompose the incentives for learning and choosing aligns the machine learner \textit{explicitly} in theory. 

% Alternatively interpreted, \autoref{thm:alignment-outputting} provides an analytical adjustment formula \eqref{eq:prediction-solution-conditional}, which can be used to modify prediction outputs ex post. 
% Using this combination of interpretations, we next discuss how our incentive-based approach relates to and unifies the existing literature on cost-sensitive learning and the adjacent problem of class imbalance.

\newpage

\section{Proofs}

\begin{proof}[Proof of \autoref{thm:alignment-outputting}]
For ease of reference with the statement in Theorem \ref{thm:alignment-outputting}, we restrict to the classification problem $\mathcal{A} = \mathcal{Y}$. 
However, note that the same result and argument hold for downstream choices and preferences in any finite set $\mathcal{A}$.

Nondegeneracy \eqref{eq:nondegen-cost} of $u$ implies that, for any class distribution $q \in \Delta (\mathcal{Y})$, the expected utility is positive for some label $y'$, so that the sum across labels is positive:
\[
    \sum_{y' \in \mathcal{Y}} \bar{\cl}(y',q) > 0
\]
Therefore we can also define the normalized vector of expected utilities $\bar{h}^\cl (q) \in \Delta (\mathcal{Y})$ as: 
\begin{equation}
    \label{eq:h-bar}
    \bar{h}_y^\cl (q) \equiv \cfrac{ \bar{\cl} (y, q) }{\sum_{y' \in \mathcal{Y}} \bar{\cl} (y', q) }
\end{equation}

Next, recall the prediction problem \eqref{eq:prediction-conditional}. Decomposing cost-weighted prediction loss and rearranging terms yields: 
\begin{align*}
    \bar{\pl}^\cl (p, q) 
    &=
    \sum_{y \in \mathcal{Y}} \pl^\cl (p,y) q_y \\
    &=
    \sum_{y \in \mathcal{Y}} \left[ \sum_{y' \in \mathcal{Y}} \pl(p,y') \cl (y',y) \right] q_y  \\
    &= 
    \sum_{y' \in \mathcal{Y}} \pl (p,y') \sum_{y \in \mathcal{Y}} \cl (y', y) q_y \\
    &= 
    \sum_{y' \in \mathcal{Y}} \pl (p,y') \bar{\cl} (y',q) 
    = \bar{\pl} (p, \bar{\cl} (\cdot, q))
    % = \langle L(p, \cdot), \bar{C} (\cdot, q) \rangle
\end{align*}
Therefore the optimal prediction given a latent posterior $q$ solves: 
\[
    \argmin_{p \in \Delta (\mathcal{Y})} 
    \, \bar{\pl} (p, \bar{\cl} (\cdot, q))
    \quad \text{or equivalently,} \quad
    \argmin_{p \in \Delta (\mathcal{Y})} \, 
    \bar{\pl} (p, \bar{h}^\cl (q))
    % \sum_{y \in \mathcal{Y}} \pl (p,y) \times \bar{h}_y^\cl (q) 
\]
since the latter objective just entails division of the former objective by a positive constant $\sum_{y' \in \mathcal{Y}} \bar{\cl} (y', q)$. 
Since the unweighted loss function $\pl$ is strictly proper \eqref{eq:proper}, it follows that $\bar{h}^\cl (q)$ is the unique optimal solution to the prediction problem \eqref{eq:prediction-conditional}. 
The closed-form solution \eqref{eq:prediction-solution-conditional} follows by definition \eqref{eq:h-bar} of $\bar{h}^u (q)$.
\end{proof}

\bigskip

\begin{proof}[Proof of \autoref{thm:incentives-learning}]
By definition of indirect learning \eqref{eq:indirect-learning} and the optimal prediction \eqref{eq:prediction-conditional},
\begin{equation}
    \label{eq:indirect-loss-min}
    \tilde{\pl}^\cl (q) = \min_{p \in \Delta (\mathcal{Y})} 
    \, \langle \pl^\cl (p, \cdot), q \rangle
    % \sum_{y \in \mathcal{Y}} \pl^\cl (p,y) \times q_y
\end{equation}
As a pointwise minimum of linear functions in $q$, $\tilde{\pl}^\cl$ is concave. 
By Theorem \ref{thm:alignment-outputting}, there is a unique minimizer $p^\cl (q)$ for each $q$, and thus a unique subgradient $\ell^\cl (p^\cl (q), \cdot)$ to $\tilde{\ell}^{\cl}$ at $q$ (e.g., Corollary 4.4.4 of \textcite{hiriart2004convex}). 
By Theorem 25.1 of \textcite{rockafellar1970convex}, $\tilde{\ell}^\cl$ is differentiable at $q$, and its derivative is the unique subgradient.
Proposition \ref{thm:incentives-learning} simply expresses this derivative component-wise.
\end{proof}

\bigskip

\begin{proof}[Proof of \autoref{thm:recalibrate}]
    Throughout this proof, we alternatively represent the utility function $u: \mathcal{Y} \times \mathcal{Y} \to \mathbb{R}_+$ as an $\ny \times \ny$ matrix $U$, with rows and columns corresponding to labels and classes, respectively, and let $U^{-1}$ denote its inverse.
    Similarly, denote probability distributions $p,q \in \Delta (\mathcal{Y})$ as column vectors, and let superscript $T$ denote the transpose operator. 
    The optimal utility-weighted prediction from Theorem \ref{thm:alignment-outputting} expressed in matrix form is:
    \begin{equation}
        \label{eq:prediction-matrix}
        p^\cl (q) = \frac{1}{\iota^{T} Uq } Uq
    \end{equation}
    where $\iota$ is defined as a column vector of ones and the term $\iota^{T} Uq$ is a real-valued function of $q$.

    Fix $q \in \Delta (\mathcal{Y})$.
    Since $U$ is nonnegative and nondegenerate, 
    $Uq \in \mathbb{R}_+^m \setminus \{ 0 \}$, and so there exists a unique scalar multiplier $\alpha > 0$ such that $\alpha Uq \in \Delta (\mathcal{Y})$.
    By \eqref{eq:prediction-matrix}, $\alpha = (\iota^T U q)^{-1}$ and $p^{\cl} (q) = \alpha Uq$.
    Therefore:
    \[
        U^{-1} [p^{\cl} (q)] = U^{-1} [\alpha Uq] = \alpha U^{-1}U  q = \alpha q
    \]
    Dividing the first and last terms of the preceding equalities by $\alpha > 0$,
    \[
    \alpha^{-1} U^{-1} [ p^u (q)] = q
    \]
    Since $\alpha > 0$ and $q \in \Delta (\mathcal{Y})$, we have $U^{-1} [ p^\cl (q) ] \in \mathbb{R}_+^n \setminus \{ 0 \}$, and it must be that $\alpha = \iota^T U^{-1} p^\cl (q)$. Thus, the desired inverse construction: 
    \[
        [p^\cl]^{-1} (p) = \frac{U^{-1} p}{\iota^T U^{-1} p}
    \]
    follows.
    Again, equivalence with \eqref{eq:calibration} is easily verified. 
\end{proof}

\begin{proof}[Proof of \autoref{thm:convex-closed}]
For what follows, recall our matrix notation in \eqref{eq:prediction-matrix} from \autoref{thm:recalibrate}:
\[
    p^\cl (q) = \frac{1}{\iota^{T} Uq } Uq
\]
Our first claim is that for any $q, q' \in \Delta (\mathcal{Y})$ such that $Uq = \beta Uq'$ for some $\beta > 0$, we have $p^\cl (q) = p^\cl (q')$. This follows because:
\[
    p^\cl (q) = \frac{1}{\iota^{T} Uq } Uq = \frac{1}{ \iota^T \beta Uq'} \beta Uq' = \frac{1}{ \iota^T Uq'} Uq' = p^\cl (q')
\] 
Our second, converse claim is that for any two posteriors $q, q' \in \Delta (\mathcal{Y})$ that generate the same analytical prediction $p^\cl (q) = p^\cl (q') = p$, it must be that $Uq = \beta Uq'$ for some $\beta > 0$. Suppose not. Then there exist realizations $y,y'$ and $\gamma > 0$ such that $(Uq)_y = \gamma (Uq)_{y'}$, but $(Uq')_y \neq \gamma (Uq')_{y'}$. 
Yet, the respective equality and inequality are preserved under any scalar division, implying that $p^\cl (q) \neq p^\cl (q')$, a contradiction to our premise.

Having established by the second claim that $Uq = \beta Uq'$ for some $\beta > 0$, it follows for any $\alpha \in [0,1]$ that:
\begin{align*}
U [ \alpha q + (1-\alpha)q'] =
\alpha Uq + (1-\alpha) Uq' = 
[ \alpha \beta + (1-\alpha)] Uq' 
\end{align*}
The desired result then follows by the first claim.
\end{proof}
\begin{proof}[Proof of \autoref{thm:decomposition}]
The proof makes use of an indicator function (i.e., a one-hot encoder) $\delta: \mathcal{Y} \to \{0,1\}^m$ defined by $\delta_{y} (y') \equiv I \{ y' = y \}$.
\begin{align*}
    \mathbb{E} [ \pl (P, Y)] 
    &= 
    \mathbb{E} \left[ 
    \langle \pl (P, \cdot), \delta (Y) \rangle
    % \sum_{y \in \mathcal{Y}}\pl(P, y) \delta_y (Y) 
    \right] \\
    &= 
    \mathbb{E} \left[ \left. \mathbb{E} \left[ 
    \langle \pl (P, \cdot), \delta (Y) \rangle
    % \sum_{y \in \mathcal{Y}} \pl(P, y) \delta_y (Y) 
    \right| P \right] \right] \\
    &= 
    \mathbb{E} \left[ 
    \langle \pl (P, \cdot ), \mathbb{E} \left[ \delta (Y) | P \right] \rangle
    % \sum_{y \in \mathcal{Y}} \pl(h(X), y) \mathbb{E} \left[ \delta_y (Y) | h(X) \right] 
    \right] \\
    &= 
    \mathbb{E} \left[ 
    \langle \pl (P, \cdot), Q \rangle
    % \sum_{y \in \mathcal{Y}} \pl(h(X), y) Q_y^h 
    \right] \\
    &= 
    \mathbb{E} [ \bar{\ell} (P, Q) ]
    % \mathbb{E} \left[ \mathbf{\pl}(f(X)) \cdot Q^f \right]
\end{align*}
\end{proof}

\section{Figures}
\label{sec:figs}

In the figures that follow, we disaggregate the loss in the test sample across five training runs per training incentive. 
We conduct multiple training runs in order to disentangle systematic changes in performance from stochasticity in the training procedure.
For each run, the point denotes the optimal performance and the training step at which it is achieved. 

\begin{figure}[H]
    \centering
    \includegraphics[width=\linewidth]{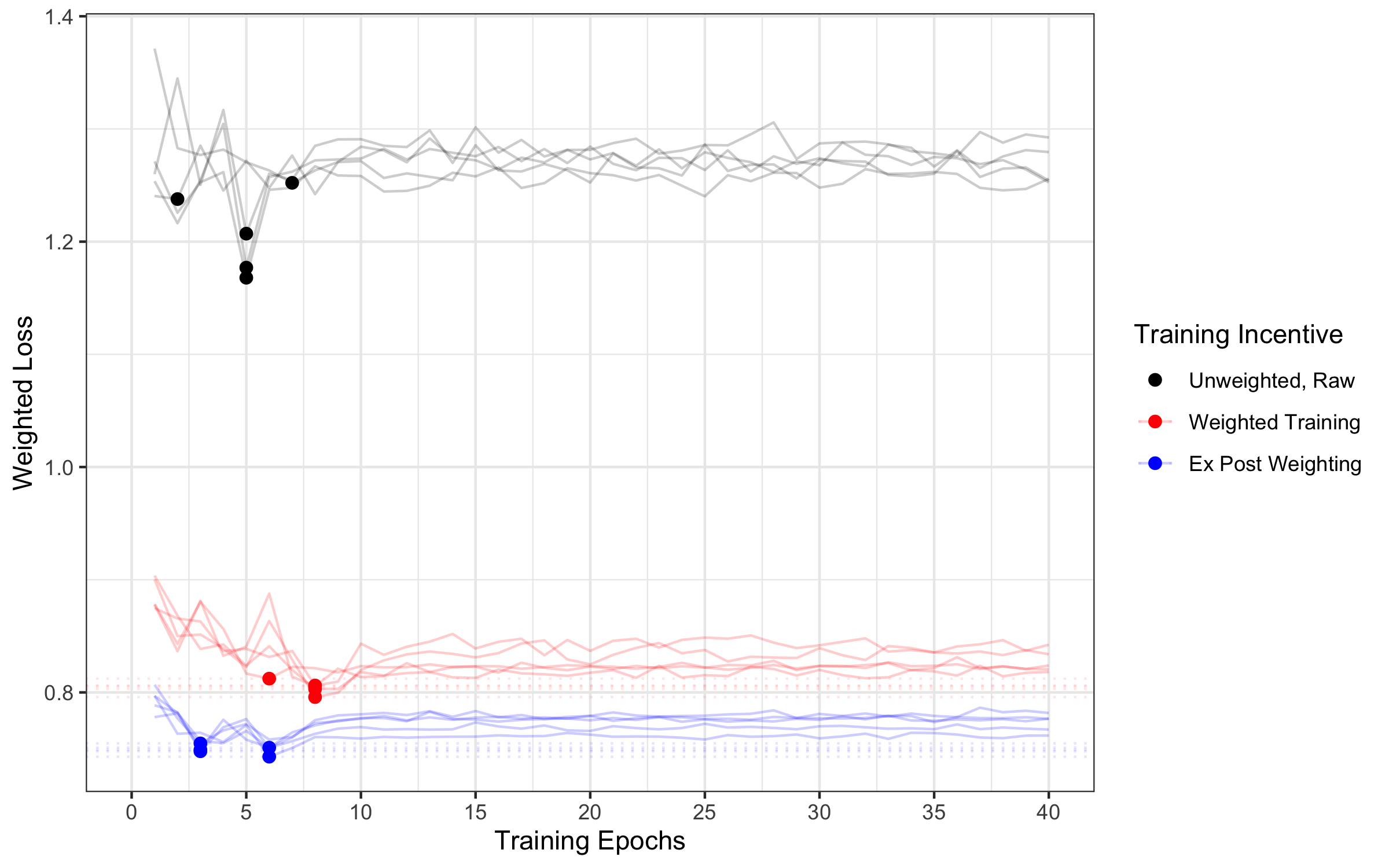}
    \caption{Weighted loss when emphasizing pneumonia, evaluated in test sample. The black lines represent the weighted loss (across training runs) from unweighted training without adjusting predictions. The red lines represent the weighted loss from weighted training. 
    Finally, the blue lines represent the weighted loss from unweighted training after analytically adjusting predictions. 
    For each run, the point denotes the minimal weighted loss and the training epoch at which it is achieved. 
    Consistently across training runs, we outperform the machine on its own objective by not training according to downstream incentives, but rather analytically adjusting for them ex post.}
    \label{fig:loss-pneumonia}
\end{figure}

For our leading case of pneumonia instances in chest X-rays, we present figures extended in two further ways. 
First, when plotting performance according to weighted loss (Figure \ref{fig:loss-pneumonia}), we include the \textit{uncorrected} predictions from unweighted training [\textit{Unweighted, Raw}, in black]. 
As is to be expected, these perform worse according to weighted loss than training according to that objective itself. 
However, this conflates an issue of misspecified predictions with their underlying information content. 
By analytically correcting the unweighted predictions [\textit{Ex Post Weighting}, in blue], we isolate the information channel and achieve better performance on weighted loss than training on that objective [\textit{Weighted Training}, in red].
Across five runs, we achieve better performance on the weighted objective under the \textit{Ex Post Weighting} procedure than under \textit{Weighted Training}, with a mean improvement of 6.9\%.

\begin{figure}[H]
    \centering
    \includegraphics[width=\linewidth]{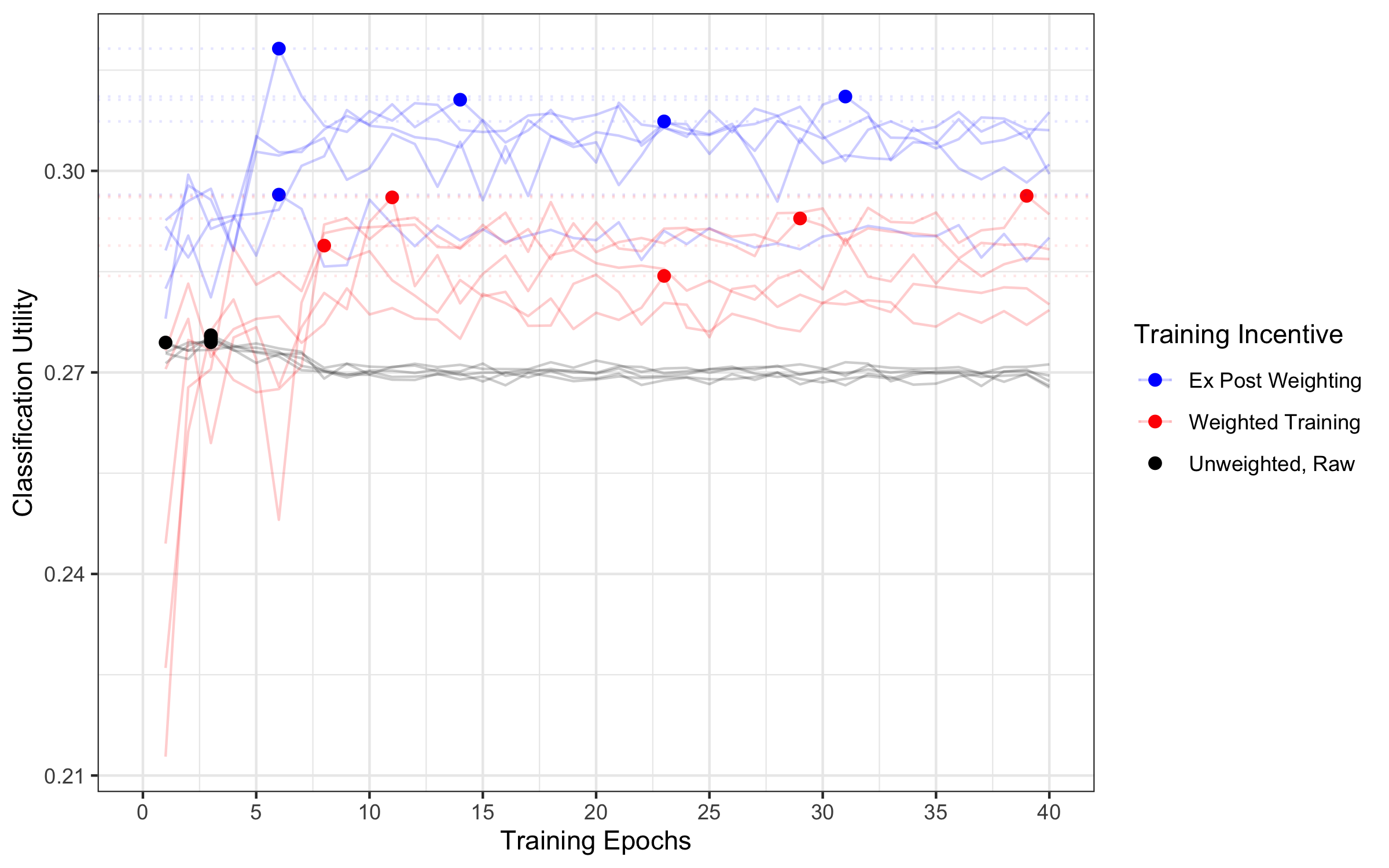}
    \caption{Classification utility when emphasizing pneumonia, evaluated in test sample. The black lines represent the achieved utility (across training runs) from unweighted training without adjusting predictions. 
    The red lines represent utility from weighted training. 
    Finally, the blue lines represent utility from unweighted training after analytically adjusting predictions. 
    For each run, the point denotes the maximal classification utility and the training epoch at which it is achieved. 
    Consistently across training runs, we outperform the machine on the downstream utility objective by not training according to downstream incentives, but rather analytically adjusting for them ex post.
    Based on the preceding Figure \ref{fig:loss-pneumonia}, we attribute this to underperformance and suppressed learning when the machine is trained according to utility-weighted cross entropy.
    }    \label{fig:utility-pneumonia}
\end{figure}

Second, we include an analogous figure of classification utility (Figure \ref{fig:utility-pneumonia}). 
Perhaps surprisingly, the training epoch that minimizes weighted loss is not necessarily the one that also maximizes classification utility. 
We attribute this to the discontinuity of classification utility in predictions, combined with 
a heavy dependence on single observations of a rare but heavily weighted class.
Nevertheless, our main point is robust: training under the unweighted objective and analytically correcting predictions consistently outperforms training under the weighted objective, even according to that objective. 
Across five runs, the worst performance of the unweighted procedure is approximately equal to the best performance of the weighted procedure (achieving in each case a classification utility of 0.296). 

% Note: NOT accuracy. Utility is \textit{normalized} weights as detailed in \eqref{eq:cost-normal}.

\begin{figure}[H]
    \centering
    \includegraphics[width=\linewidth]{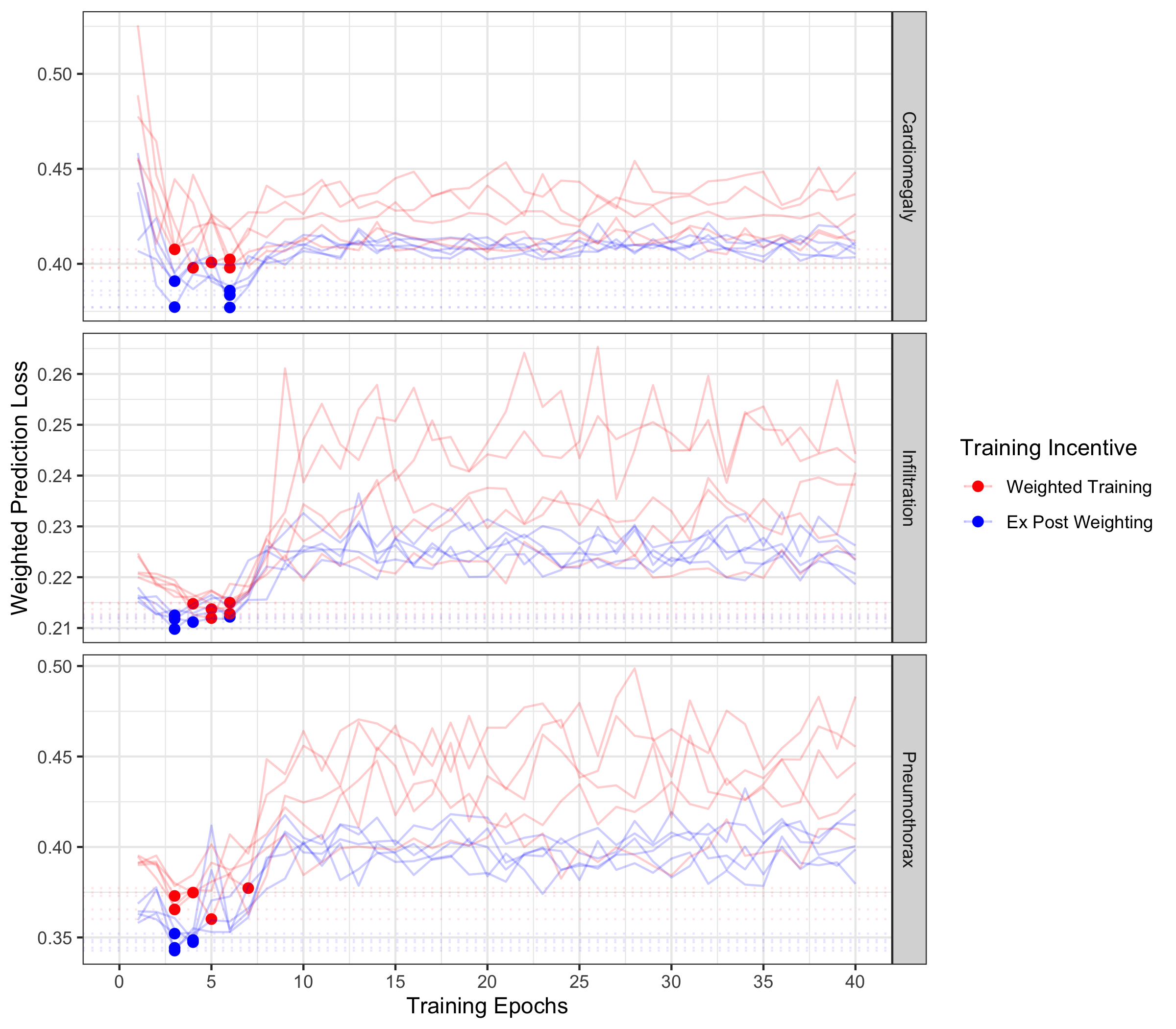}
    \caption{Weighted loss when respectively emphasizing cardiomegaly, infiltration, and pneumothorax, evaluated in test sample. 
    The red lines represent the weighted loss from weighted training across five training runs. 
    The blue lines represent the weighted loss from unweighted training after analytically adjusting predictions. 
    For each run, the point denotes the minimal weighted loss and the training epoch at which it is achieved. 
    On average and (with the exception of infiltration) consistently across training runs, we outperform the machine on its own objective by not training according to downstream incentives, but rather analytically adjusting for them ex post.}
    \label{fig:loss-other}
\end{figure}

\begin{figure}[H]
    \centering
    \includegraphics[width=\linewidth]{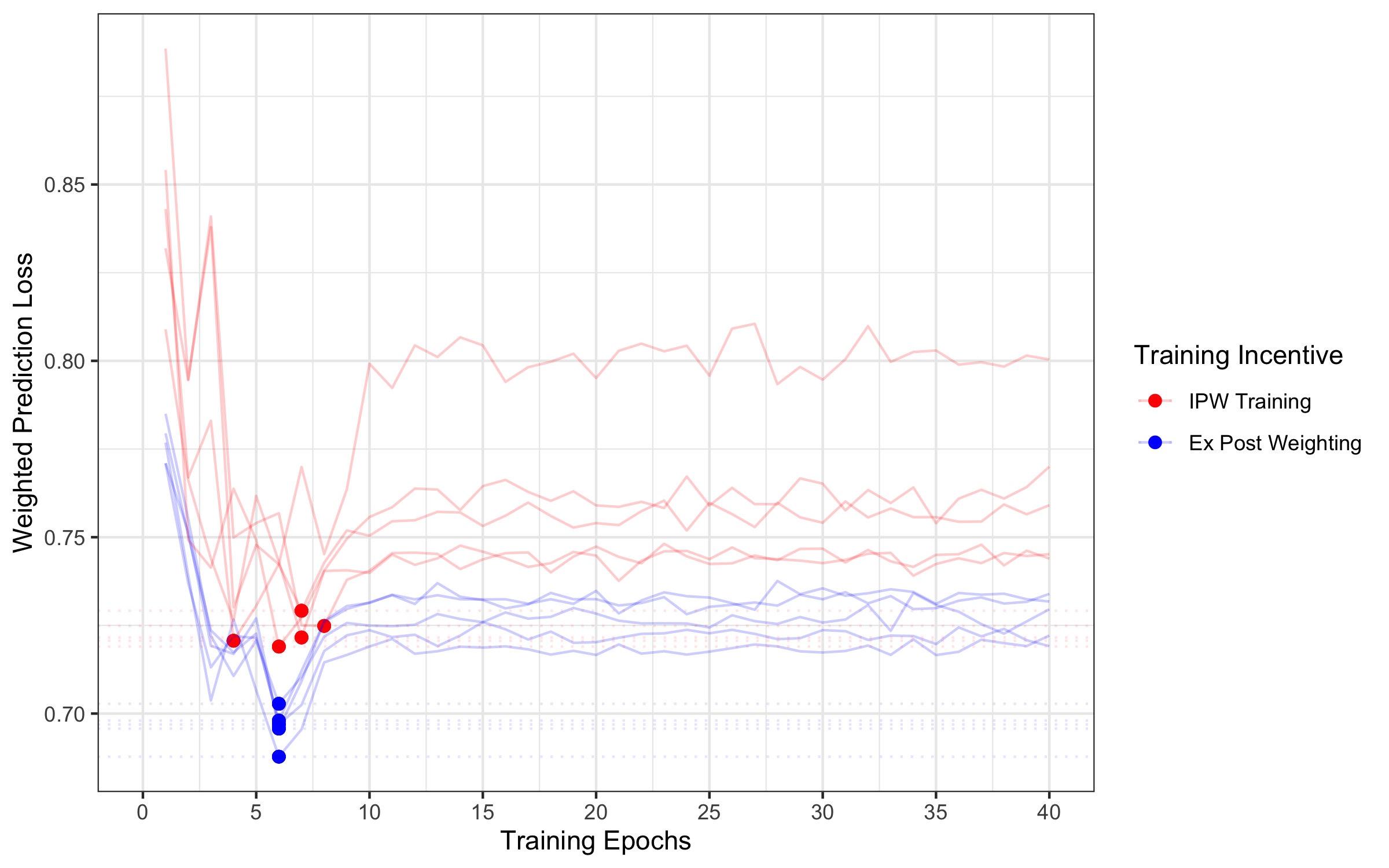}
    \caption{Weighted loss when balancing the data through inverse probability weighting. 
    The red lines represent the weighted loss from weighted training across five training runs. 
    The blue lines represent the weighted loss from unweighted training after analytically adjusting predictions. 
    For each run, the point denotes the maximal classification utility and the training epoch at which it is achieved. 
    Consistently across training runs, we outperform the machine on its own objective by not training with inverse probability weights, but rather analytically adjusting predictions to account for the weights ex post.}
    \label{fig:loss-ipw}
\end{figure}

\begin{figure}[H]
    \centering
    \includegraphics[width=\linewidth]{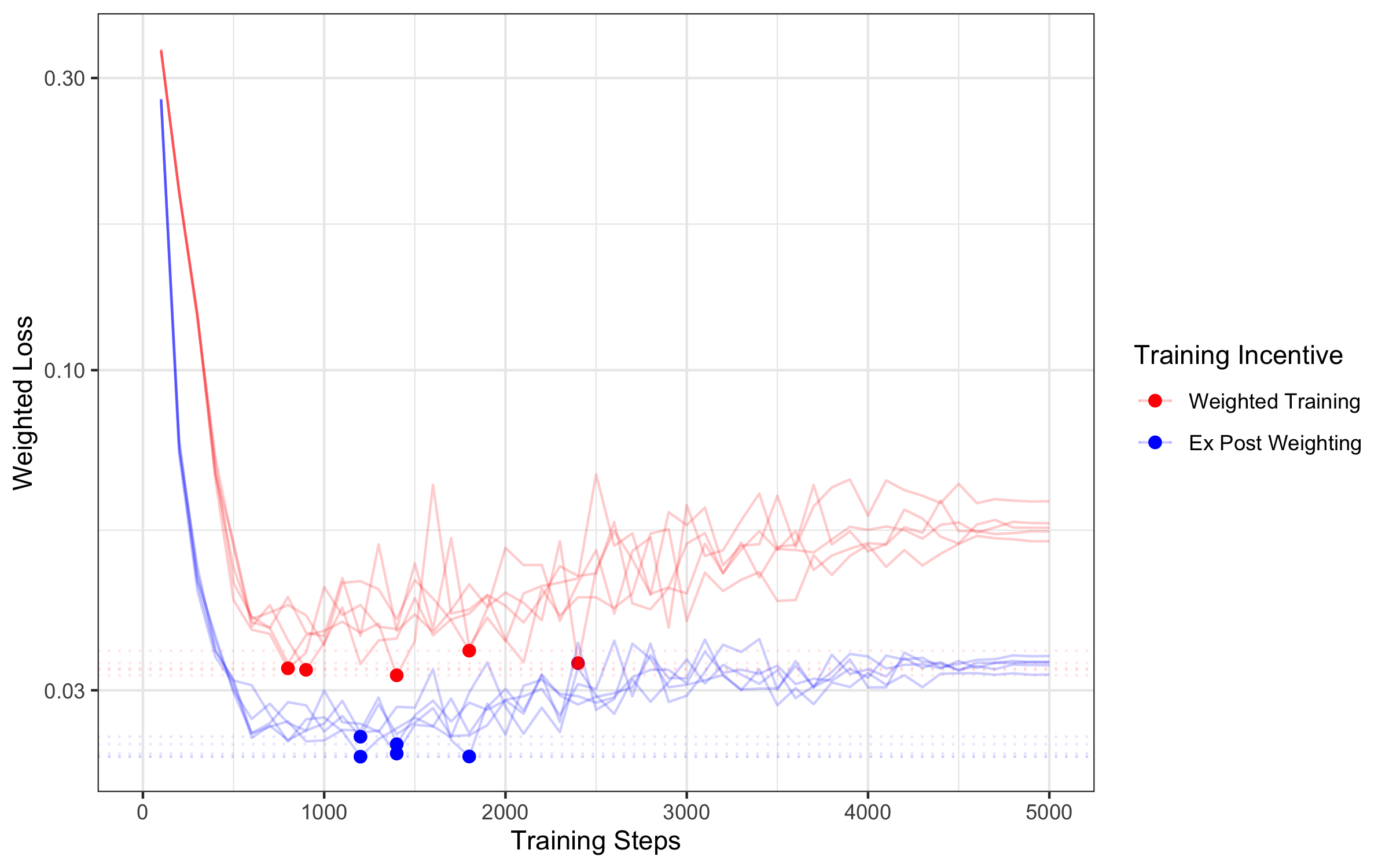}
    \caption{Weighted loss when emphasizing the (most difficult) class ``cat'' in CIFAR-10 data, evaluated in the test sample. 
    The red lines represent the weighted loss from weighted training across five training runs. 
    The blue lines represent the weighted loss from unweighted training after analytically adjusting predictions. 
    For each run, the point denotes the minimal weighted loss and the training step at which it is achieved. 
    Consistently across training runs, we outperform the machine on its own objective by not training according to downstream incentives, but rather analytically adjusting for them ex post.}    
    \label{fig:loss-cifar10}
\end{figure}

\begin{figure}[H]
    \centering
    \includegraphics[width=\linewidth]{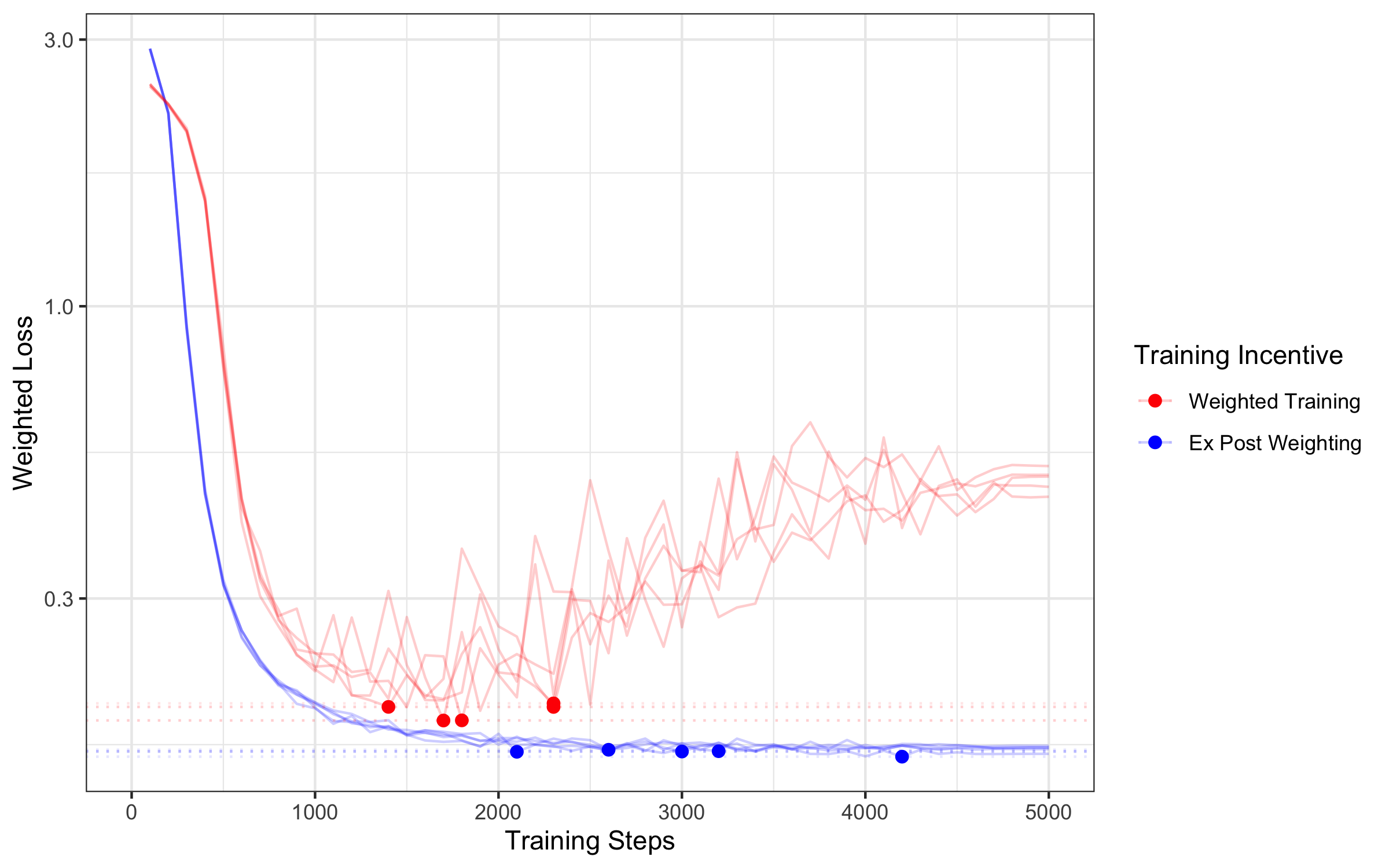}
    \caption{Weighted loss when emphasizing the (most difficult) class ``maple tree'' in CIFAR-100 data, evaluated in the test sample. 
    The red lines represent the weighted loss from weighted training across five training runs. 
    The blue lines represent the weighted loss from unweighted training after analytically adjusting predictions. 
    For each run, the point denotes the minimal weighted loss and the training step at which it is achieved. 
    Consistently across training runs, we outperform the machine on its own objective by not training according to downstream incentives, but rather analytically adjusting for them ex post.}       
    \label{fig:loss-cifar100}
\end{figure}

\section{Data and Training Procedures}
\label{sec:data}
\subsection{Chest X-Rays}

Our model training procedure follows that of the CheXNeXt algorithm (\cite{rajpurkar2018deep}), in which a deep neural network was trained using the ChestX-ray14 dataset of \textcite{wang2017chestxray}.
The ChestX-ray14 dataset consists of 112,120 frontal chest X-rays that were synthetically labeled with up to fourteen thoracic diseases. 
Our code for model training is adapted from the publicly available CheXNeXt codebase of \textcite{rajpurkar2018deep}. 

As in their work, we adopt random horizontal flipping and normalize based on the mean and standard deviation of images in the ImageNet dataset (\cite{deng2009imagenet}). 
For each model, we train a 121-layer dense convolutional neural network (DenseNet, \cite{huang2017densenet}) with network weights initialized to those pretrained on ImageNet, using Adam with standard parameters 0.9 and 0.999 (\cite{kingma2015adam}), and using batch normalization (\cite{ioffe2015batch}). 
We use an initial learning rate of 0.0001 that is decayed by a factor of 10 each time the validation loss plateaus after an epoch. 

Besides our class-weighting modification to the loss function, we modify their implementation in three ways.
First, we restrict attention to the subset of 91,324 images with a single label (including ``No Finding'' of disease) and train multi-class classifiers using a 70-20-10 training-test-validation split. 
In related work, \textcite{caplin2022modeling} focus on binary classification of pneumonia and find statistically significant but economically small effects consistent with our results; this suggests that the extent of misalignment increases with the complexity of the problem and prediction space.
Second, rather than conduct early stopping based on validation loss, we run each instance for 40 training epochs to compare the evolution of loss in the test sample and show that it is pointwise ordered across our training regimes. (The validation loss is still used implicitly by the algorithm to update the learning rate.)
Third, we trade off a higher batch size of 16 at the expense of a slightly smaller imaging scaling size of 224 by 224 pixels (instead of a batch size of 8 and an image rescaling of 512 by 512 pixels, respectively). 

\subsection{CIFAR}
We fine-tune and evaluate the Base variant of the Vision Transformer model with 16x16 pixel patch size (ViT-B16, \cite{dosovitskiy2021image}), pre-trained on the ImageNet-21k dataset (\cite{deng2009imagenet}), on the CIFAR-10 and CIFAR-100 datasets (\cite{krizhevsky2009cifar}).
Summarizing their training implementation, the Base model contains 12 layers, hidden size 768, MLP size 3072, 12 heads, and 86 million parameters. 
The model is trained using Adam (\cite{kingma2015adam}) with $\beta_1 = 0.9, \beta_2 = 0.999$, a batch size of 4,096, and a high weight decay of 0.1. 
In the fine-tuning stage, we follow their implementation with a batch size of 512, linear learning rate warm-up and cosine decay with a base learning rate of 0.03, stochastic gradient descent with momentum of 0.9, and gradient clipping at global norm of 1. 
The only modification we make from their implementation is to vary class weighting in the loss function.

\end{document}